\documentclass{article}
\usepackage{graphicx} % Required for inserting images
\usepackage[numbers,sort&compress]{natbib}
\usepackage{authblk}
\usepackage{subcaption}
\usepackage{hyperref}
\usepackage{float}
\usepackage{booktabs}
\usepackage{amsmath}
\usepackage{xcolor} % Required for changing text color to grey
\usepackage{etoolbox}

\AtBeginDocument{%
  \apptocmd{\thebibliography}{%
    \scriptsize
    \setlength{\itemsep}{0pt}%
    \setlength{\parsep}{0pt}%
    \setlength{\parskip}{0pt}%
  }{}{}%
}

\usepackage[margin=2.5cm, top=1.5cm, headheight=10pt, headsep=15pt]{geometry} 

\usepackage{fancyhdr}
\title{A Bayesian Optimized Biquad Filter-Burg Model for Real-Time EEG Phase Estimation}
\author[1,3]{Ivan Martinez-Kay}
\author[1,4]{Tatiana Kardashina}
\author[2,3]{ShiNung Ching}
\author[1,4]{Eric C. Leuthardt}
\author[1]{Benjamin T. Acland}

\affil[ ]{Washington University in St.\ Louis School of Medicine}
\affil[1]{The Taylor Family Department of Neurosurgery}
\affil[2]{Department of Neurology}
\affil[ ]{Washington University in St.\ Louis, The McKelvey School of Engineering\\\rule{0pt}{1.5em}}
\affil[3]{The Preston M. Green Department of Electrical and Systems Engineering}
\affil[4]{Department of Mechanical Engineering and Materials Science}

\date{November 3, 2026}

\begin{document}

\maketitle

\begin{abstract}
    Real-time phase estimation on stimulus-aligned transient EEG is critical for closed-loop neuromodulation systems targeting the theta-band, yet most real-time phase estimation methods have been developed and validated on resting-state, high-fidelity signals in the alpha band. Recent work has pursued deep learning architectures, while individualized classical bandpass filter tuning remains underexplored.

    This study characterizes the spectral and temporal properties of non-anticipative bandpass filtering and demonstrates the performance gains achieved through individualized tuning. We present a Bayesian-Optimized Biquad Filter-Burg (BOB) model evaluated on EEG recorded during peripheral vibrotactile stimulation of the left median nerve (theta, N=20 healthy subjects) and on resting-state recordings from the Max Planck Institute’s LEMON dataset (alpha, N=20 healthy subjects). BOB was benchmarked against the Deep Dual-Branch Complex Neural Network (DBCNN), EEG Phase Prediction Network (EPN), Phastimate, and Phase Estimation with Autoregressive Padding (PEAP) under matched testing conditions: channel selection, individualized optimization, evaluation time points, and an offline reference phase obtained through zero-phase filtering and the Hilbert transform. Under these testing conditions, BOB achieved matched or better performance than these models at a fraction of the calibration time.
\end{abstract}

\section{Introduction}

Closed-loop neuromodulation systems \citep{zrenner2016twoloops} that deliver stimuli contingent on the instantaneous phase of ongoing EEG oscillations have emerged as a powerful tool for probing cortical excitability \citep{zrenner2018realtime, chen2013realtime, mansouri2017fast}. By timing stimulation to specific phases of a target rhythm, these frameworks enable researchers to test whether phase-dependent differences in neural responsiveness influence sensory, plasticity-related, or motor responses \citep{zrenner2018realtime}. The clinical promise of this approach depends on the technical requirement of estimating the instantaneous phase of a target oscillation in real time, and with sufficient accuracy to meaningfully constrain stimulation timing \citep{zrenner2023rising}.

The standard approach, autoregressive (AR) forward prediction, fits an AR model to a window of bandpass-filtered EEG, extrapolates the signal forward by the system's processing latency, and extracts phase via the Hilbert transform \citep{chen2013realtime, zrenner2018realtime}. Chen et al.\ \citep{chen2013realtime} proposed this framework on intracranial theta oscillations using Yule--Walker coefficient estimation and a genetic algorithm for parameter optimization. Their pipeline applies a zero-phase (bidirectional) bandpass filter on past and future samples extrapolated by the model to remove group delay. Subsequent scalp-EEG implementations adopted real-time filtering but carried several underappreciated parameters that are typically fixed to heuristic defaults or left unoptimized on a per subject basis \citep{shirinpour2020etp, zrenner2020shaky}. Shirinpour et al.\ explicitly noted that AR parameter optimization lacks any principled method, and that filter type and order can strongly affect performance \cite{shirinpour2020etp}. Zrenner et al.\ further demonstrated that phase estimation accuracy is fundamentally bounded by signal-to-noise ratio (SNR) and that the ground truth used to evaluate real-time methods is itself subject to estimation error \citep{zrenner2020shaky, kim2023eeg}.

To bypass AR parameter sensitivity, Shirinpour et al.\ \citep{shirinpour2020etp} initially proposed the Educated Temporal Prediction (ETP) algorithm, which learns the mean inter-peak interval of a target oscillation from a recent baseline segment and predicts subsequent peak times by extrapolating that interval from the most recently detected peak, without fitting an AR model. ETP was shown to outperform both Fourier-based methods and Yule--Walker extrapolation on resting alpha band data, while remaining parameter-free and computationally lightweight. However, ETP's reliance on peak detection requires a stable, continuous, high-SNR oscillation, a condition that is not met by transient, event-related rhythms in the theta-band. Shirinpour et al.\ \citep{shirinpour2025btp} subsequently reported that ETP failed to track theta oscillations in any participant tested under real-time conditions. Their follow-up algorithm, Bayesian Temporal Prediction (BTP), extended ETP by using Gaussian process based Bayesian optimization \citep{brochu2010tutorial} to identify personalized prediction parameters \citep{figee2021personalized}, demonstrating improved accuracy across multiple frequency bands and cortical regions. BTP established that per-subject parameter learning substantially improves phase prediction, but addresses the task differently by predicting when a target phase will next occur, rather than estimating the instantaneous phase at an arbitrary evaluation point.

Adjacent to this work, machine learning approaches have sought to replace the signal processing chain entirely. McIntosh and Sajda \cite{mcintosh2020mlof} proposed the optimized multi-layer filter (MLOF) architecture, which uses supervised learning to approximate the mapping from real-time, minimally processed EEG to the phase estimate produced by a zero-phase filter. While MLOF outperformed standard real-time FIR filter approaches on alpha-band data, a gated recurrent unit (GRU) architecture offered no further improvement \citep{mcintosh2020mlof, zeng2023transformers}. Liu et al.\ \citep{liu2024epn} proposed the EEG Phase Prediction Network (EPN), a neural network that maps raw broadband EEG directly to instantaneous phase. Evaluated on their own Transcranial Magnetic Stimulation (TMS) alpha-band data, EPN achieved $77.3\%$ phase accuracy in an offline experiment and outperformed Yule--Walker, MLOF, and ETP. More recently, Hamdan et al.\ \citep{DBCNNHamdan} proposed a deep dual-branch complex neural network (DBCNN) for instantaneous phase estimation, validated primarily on hippocampal theta ($6$--$9$~Hz) local field potentials recorded from 12 rats during spatial navigation. The authors also demonstrated real-time FPGA deployment in a freely moving rat and included a proof-of-concept evaluation on human alpha EEG. The performance of these deep learning approaches on transient, stimulus-locked rhythms remains unknown. 

Stimulus-onset theta-band phase estimation presents a new challenge within this literature. Unlike the high-SNR resting-state alpha recordings used in our comparison, the stimulus-aligned theta recordings have a less distinct oscillatory peak and are evaluated during the early post-stimulus response. Furthermore, theta is typically low in amplitude, transient, and strongly entrained by external events rather than present at rest \citep{gordon2021prefrontal, vanengen2026dissociating, donoghue2020fooof}, making the task plausibly harder. Gordon et al.\ \citep{gordon2021prefrontal} demonstrated that real-time theta-phase TMS is feasible using AR prediction with adapted spatial filters and a power-gating strategy, but noted substantially higher estimation error relative to alpha-band applications and the need for individualized processing pipelines. Critically, it remains unclear how joint per-subject tuning of the filter--AR signal processing chain balances spectral selectivity and temporal alignment, and whether deep learning architectures offer advantages over a well-optimized classical approach in this regime.

This study addresses this gap directly and presents a per-subject Bayesian optimization framework that jointly tunes five parameters of a biquad filter-Burg model using a Gaussian process surrogate on a short calibration split. This framework is evaluated on IRB-approved EEG recorded during peripheral vibrotactile stimulation of the median nerves and open-source resting-state alpha rhythm EEG from the Max Planck Institut Leipzig Mind-Brain-Body Dataset - LEMON \citep{babayan2019mind}.

\section{Methods}

\subsection*{Experimental Paradigm}

For evoked theta evaluation, EEG was recorded from 20 healthy subjects using a DSI-24 wireless dry electrode system with 24 channels in the international 10-20 montage, sampled at 300~Hz. The vibrations were delivered by eccentric rotating mass motors to the median nerves of the left hand at frequencies 2, 6, 12, 20, and 40 Hz, with 5-second stimulation periods separated by rest intervals of 5, 6, or 7~s (selected at random).

\subsubsection*{EEG Preprocessing} 

Because our objective was maximal theta SNR for benchmarking phase estimation rather than source localization, we selected a single optimal channel per subject based on the calibration split rather than imposing a spatial montage. Channel selection was performed on preprocessed data (artifact rejection, ICA, and a 0.5–40~Hz band-pass) to identify the channel carrying theta with the highest SNR on a per subject basis. Cz, C3, F3, and Fz were the consistent candidates across subjects, and the channel with the highest theta SNR among these was selected for each subject.  We note that a single referential channel does not disambiguate the underlying generator and that the selected channel may differ across subjects; because phase estimation does not require source-level interpretation, we judged this trade-off acceptable.

Stimulus event markers were then extracted from the raw BCI2000 files.
For each subject, 125 epochs were defined from $-1$~s to $+3$~s relative
to stimulus onset, yielding $1201$ samples per trial at $300$~Hz with
both endpoints included. Extended epochs spanned $-3.03$~s to $+4$~s
for PEAP ($2110$ samples) and $-2$~s to $+3$~s for DBCNN
($1501$ samples) to accommodate methodological input requirements.

\subsubsection*{LEMON dataset split}

Sensorimotor recordings were drawn from the resting-state EEG of the Leipzig Study for Mind--Body--Emotion Interactions (LEMON) dataset \citep{babayan2019mind}, downloaded as raw BrainVision (BIDS) recordings from the public FCP-INDI repository. LEMON resting-state EEG was acquired with a 62-channel active-electrode ActiCAP system in the 10--10 layout, referenced to FCz and digitized at 2500\,Hz, over interleaved eyes-open and eyes-closed blocks. Because the sensorimotor analysis targeted the alpha rhythm, subjects were pre-screened by alpha-band SNR, and the 20 subjects with the highest alpha SNR (of 215 screened) were retained.

For each retained subject, a single sensorimotor signal was derived as a C3 surround (Hjorth) Laplacian \citep{mcfarland1997spatial}. The native 2500\,Hz recordings were resampled to 1\,kHz to match PEAP's methodological constraint in filter configuration \cite{kirchhoff2026phaseestimationautoregressivepadding}, and the continuous Laplacian time series was segmented into 200--218 consecutive, non-overlapping 5\,s epochs (5000 samples each), discarding the final partial segment.

\subsubsection*{Evaluation Procedure}

Within each subject, epochs were partitioned in recording order into a calibration set (first 60\% of epochs) and a held-out test set (remaining 40\%). Epochs were neither shuffled nor pooled across subjects. Splitting chronologically rather than at random ensures that temporally adjacent epochs are not divided across the calibration/test boundary, so that test performance is not inflated by leakage between neighboring segments.

For each test epoch, the pipeline was run at each evaluation time-point, and its phase estimate was compared against an offline reference phase. The offline reference was obtained by zero-phase bidirectional filtering (\texttt{filtfilt}~\citep{Gustafsson1996}) using Hamming-windowed FIR filters of order 174 (theta ($\theta$)) and 230 (alpha ($\alpha$)) followed by the Hilbert transform \cite{marple1999computing}.

Alpha phase estimation was evaluated at 1.5, 2.0, 2.5, 3.0, and 3.5\,s within the 5\,s epoch. Theta phase prediction was evaluated at five post-stimulus time-points, 300, 400, 500, 600, and 700\,ms, to span an interval of the early onset stimulation.

\subsubsection*{Statistical Analysis}

Participants were the independent units of analysis, with trials and
evaluation points summarized within each participant. Paired comparisons
used Wilcoxon signed-rank tests \citep{wilcoxon1945individual}, and comparisons between the independent alpha and theta cohorts used Welch tests \citep{welch1947generalization}. SNR associations used Pearson correlations with permutation tests; between-cohort correlations were
compared using Fisher-$z$ tests \citep{fisher1915frequency}. Delay associations used participant-centered observations and within-participant permutations. 

All tests were two-sided, with Holm correction \citep{holm1979simple} within families of related comparisons. Confidence intervals were obtained by resampling whole participants while preserving paired observations \citep{efron1993introduction}; ordinary 95\% intervals and Bonferroni-adjusted intervals are identified in the results. Sign tests, Spearman correlations and trimmed-mean comparisons provided sensitivity checks for the corresponding analyses. Total-order Sobol indices \citep{sobol2001global} quantified parameter contributions to variation in calibration error, including interactions, over the sampled parameter distributions; parameter dominance was assessed using paired
participant-level index contrasts.

\newpage

\subsection*{BOB}
The final iteration of BOB was developed from the ablation study
described below, in which filter order and passband cutoffs were
included in the search before retaining a biquad parameterized by
center frequency and decay time.

BOB uses Optuna's \texttt{GPSampler}\citep{akiba2019optuna} (random seed 24) with 20 exploratory searches and 20 objective searches. The objective function is set to the following:\\

\begin{equation}
\min_{p,\,C,\,f_c,\,\tau,\,P}
\mathrm{MACE}_{\mathrm{cal}}
\left(p,C,f_c,\tau,P\right).
\label{eq:mace_cal}
\end{equation}

\begin{equation}
\min_{p,\,C,\,f_c,\,\tau,\,P}
\frac{1}{K}
\sum_{i=1}^{K}
\left|
\operatorname{Arg}
\left(
e^{j(\hat{\theta}_i-\theta_i)}
\right)
\right|.
\label{eq:optimization_objective}
\end{equation}

where $p$ is the AR model order; $C$ is the context duration in seconds; $f_c$ is the nominal center frequency of the biquad filter; $\tau$ is the exponential decay time constant of its response; and $P$ is the number of autoregressively predicted samples appended for padding and the search space is defined as:

\begin{table}[H]
\centering
\caption{BOB's Optimized Parameters: Per-Subject Search Ranges.}
\label{tab:bob-params}
\begin{tabular}{lcc}
\hline
\textbf{Parameter}  & \textbf{Search Range($\theta$)}  & \textbf{Search Range($\alpha$)} \\
\hline
$p$  & $[24, 100]$    & $[48, 180]$   \\
$C$ (s)    & $[0.8, 1.6]$   & $[0.8, 2.0]$  \\
$f_c$ (Hz)       & $[4, 8]$   & $[9, 13]$ \\
$\tau$ (ms)  & $[20, 650]$ & $[35, 120]$ \\
P (samples)  & $[30, 100]$   & $[60, 180]$   \\
\hline
\end{tabular}
\end{table}

\subsubsection*{Biquad Filter}

The estimation pipeline proceeds as follows for each evaluation point. First, the preprocessed epoch is passed through a bandpass \citep{oppenheim2010} filter applied in a single forward direction and initialized with steady-state initial conditions to reduce initialization transients.

The biquad is parameterized by a nominal center frequency
$f_c$ and an exponential decay time $\tau>0$, expressed in seconds. At sampling frequency $f_s$, these determine the pole angle and radius:

\begin{equation}
    \omega_c = \frac{2\pi f_c}{f_s},
    \qquad
    r = \exp\left(-\frac{1}{f_s\tau}\right).
\end{equation}

The filter transfer function is

\begin{equation}
    H(z)
    =
    g\,
    \frac{1-z^{-2}}
    {1-2r\cos(\omega_c)z^{-1}+r^2z^{-2}},
\end{equation}

where $z$ is the transform variable. The poles are located at $r e^{\pm j\omega_c}$, while the zeros at $z=\pm1$ suppress DC and the Nyquist frequency. Since $0<r<1$, the filter is stable.

The gain factor

\begin{equation}
    g =
    \frac{
        \left|1-2r\cos(\omega_c)e^{-j\omega_c}
        +r^2e^{-j2\omega_c}\right|
    }{
        \left|1-e^{-j2\omega_c}\right|
    }
\end{equation}

ensures $|H(e^{j\omega_c})|=1$. Here, $f_c$ specifies the pole frequency and gain-normalization frequency, rather than necessarily the exact frequency of maximum gain.

Letting $u[n]$ denote the preprocessed input and $x[n]$ the
filtered signal, the corresponding difference equation is

\begin{equation}
    x[n]
    =
    2r\cos(\omega_c)x[n-1]
    -r^2x[n-2]
    +g\bigl(u[n]-u[n-2]\bigr).
\end{equation}

The filter's homogeneous response contains oscillatory modes with exponentially decaying envelope

\begin{equation}
    r^n = \exp\left(-\frac{n}{f_s\tau}\right).
\end{equation}

Thus, $f_c$ sets the oscillatory pole frequency, while $\tau$ controls the persistence of the filter response.

\subsubsection*{Burg Autoregression}

At each evaluation point, the Burg algorithm~\citep{burg1975maximum} fits an autoregressive model of order $p$ to the last $N = C f_s$ samples of the filtered signal, $x[0], \ldots, x[N-1]$, where $C$ is the context duration in seconds (Table~\ref{tab:bob-params}). The algorithm recursively estimates reflection coefficients $k_m$ by minimizing the sum of forward and backward prediction error powers. At stage $m$, the reflection coefficient is

\begin{equation}
    k_m =
    \frac{
        -2\sum_{n=m}^{N-1}
        e_f^{(m-1)}[n]\,e_b^{(m-1)}[n-1]
    }{
        \sum_{n=m}^{N-1}
        \left[
            \left(e_f^{(m-1)}[n]\right)^2
            +
            \left(e_b^{(m-1)}[n-1]\right)^2
        \right]
    },
    \qquad m=1,\ldots,p.
\end{equation}

where $e_f^{(m)}$ and $e_b^{(m)}$ are the forward and backward prediction errors at stage $m$, initialized as $e_f^{(0)}[n] = e_b^{(0)}[n] = x[n]$. The errors are updated via the lattice recursion

\begin{equation}
    e_f^{(m)}[n] = e_f^{(m-1)}[n] + k_m\, e_b^{(m-1)}[n-1], \qquad
    e_b^{(m)}[n] = e_b^{(m-1)}[n-1] + k_m\, e_f^{(m-1)}[n]
\end{equation}

and the AR coefficients $\{a_k\}_{k=1}^{m}$ are updated at each stage via the Levinson recursion
\cite{levinson1947wiener, musicus1988levinson, delsarte1986split, durbin1960fitting, trench1964algorithm}. This study uses Burg estimation rather than Yule--Walker or
covariance estimators \cite{percival1993spectral, mills1990timeseries, pandit1983timeseries} to fit a stable all-pole model to short data segments \citep{kay1988modern}.

Numerical stability is enforced directly at the lattice stage. Each reflection coefficient is clipped to $\lvert k_m\rvert\leq1-\epsilon$, with $\epsilon=10^{-4}$. In exact arithmetic, reflection coefficients with magnitudes below unity ensure that the poles of the resulting AR model, defined by $1-\sum_{k=1}^{p}a_kz^{-k}=0$, lie inside the
unit circle, without explicit root computation or rescaling.

The stabilized AR model extrapolates the filtered signal forward from the evaluation point by $P$ samples (Hilbert padding). Each extrapolated sample is computed as

\begin{equation}
    \hat{x}[n] = \sum_{k=1}^{p} a_k\, x[n - k], \qquad n = N,\, N+1,\, \ldots,\, N + P - 1
\end{equation}

where $x[n-k]$ refers to either observed or previously extrapolated values. This produces a predicted signal extension that preserves the spectral characteristics of the fitted AR model.

\paragraph{Real-time zero-phase \texttt{lfilter} reconstruction:}
Let $\tilde{x} = [\,x[0], \ldots, x[N-1],\, \hat{x}[N], \ldots, \hat{x}[N+P-1]\,]$ denote the concatenation of the context window and its autoregressive extrapolation; $\tilde{x}$ is expressed in the forward-filtered (\texttt{lfilter}) domain throughout. The forward bandpass imposes a frequency-dependent group delay, so the phase read from $\tilde{x}$ is systematically lagged relative to a zero-phase target. A forward--backward filter (\texttt{filtfilt}) removes this lag but requires samples from the future of the evaluation point. The autoregressive pad circumvents this because $\hat{x}[N], \ldots, \hat{x}[N+P-1]$ are forecast from past samples alone \citep{chen2013realtime}. A second, time-reversed application of the same bandpass can be run over $\tilde{x}$ without accessing any real future sample,

\begin{equation}
    \tilde{x}_{\mathrm{zp}} \;=\; \mathcal{R}\!\left[\, \mathcal{F}_{b,a}\!\left( \mathcal{R}[\tilde{x}] \right) \right],
\end{equation}

where $\mathcal{F}_{b,a}$ is the IIR filtering operator with coefficients $(b,a)$ and $\mathcal{R}$ denotes time reversal. The forward pass and this backward pass jointly form a zero-phase filter whose group delay cancels at the read-out index $n^{*}=N-1$, aligning the estimate with the zero-phase convention of the offline reference, bypassing access to true future samples. The pad additionally supplies the samples over which the backward filter settles before reaching $n^{*}$. A single binary flag $z$ selects the mode: with $z=0$ (forward-only) phase is read from $\tilde{x}$ directly; with $z=1$ (zero-phase) $\tilde{x}$ is replaced by $\tilde{x}_{\mathrm{zp}}$. 

\paragraph{Phase extraction:}The Hilbert transform is applied to $\tilde{x}$ (replaced by $\tilde{x}_{\mathrm{zp}}$ when $z=1$), yielding the analytic signal

\begin{equation}
    z_a[n] = \tilde{x}[n] + j\, \mathcal{H}\{\tilde{x}\}[n]
\end{equation}

Instantaneous phase is read at the target index $n^* = N - 1$:

\begin{equation}
    \hat{\theta} = \mathrm{Arg}\left( z_a[n^*] \right)
\end{equation}

The role of the padding parameter $P$ is to place $n^*$ in the interior of the analytic signal rather than at its edge \cite{article, kirchhoff2026phaseestimationautoregressivepadding}, mitigating the well-known boundary distortion of the Hilbert transform~\citep{marple1999computing,article, kirchhoff2026phaseestimationautoregressivepadding}. The predicted phase is stored as $(\sin\hat{\theta},\, \cos\hat{\theta})$ to avoid discontinuities at $\pm\pi$.

\subsubsection*{Ablation Study}

To set the precedent, we adopted Blackwood et al.'s \citep{blackwood2018continuous} model using a Butterworth order 2 filter \citep{butterworth1930} with Burg AR padding providing us with a baseline for our original ablation study. To characterize the Butterworth filter's behavior across a varying search space, we observed the magnitude response, phase delay, and group delay induced by different filter orders in a non-anticipative setting. 

\begin{figure}[H]
    \centering
    \includegraphics[width=0.8\linewidth]{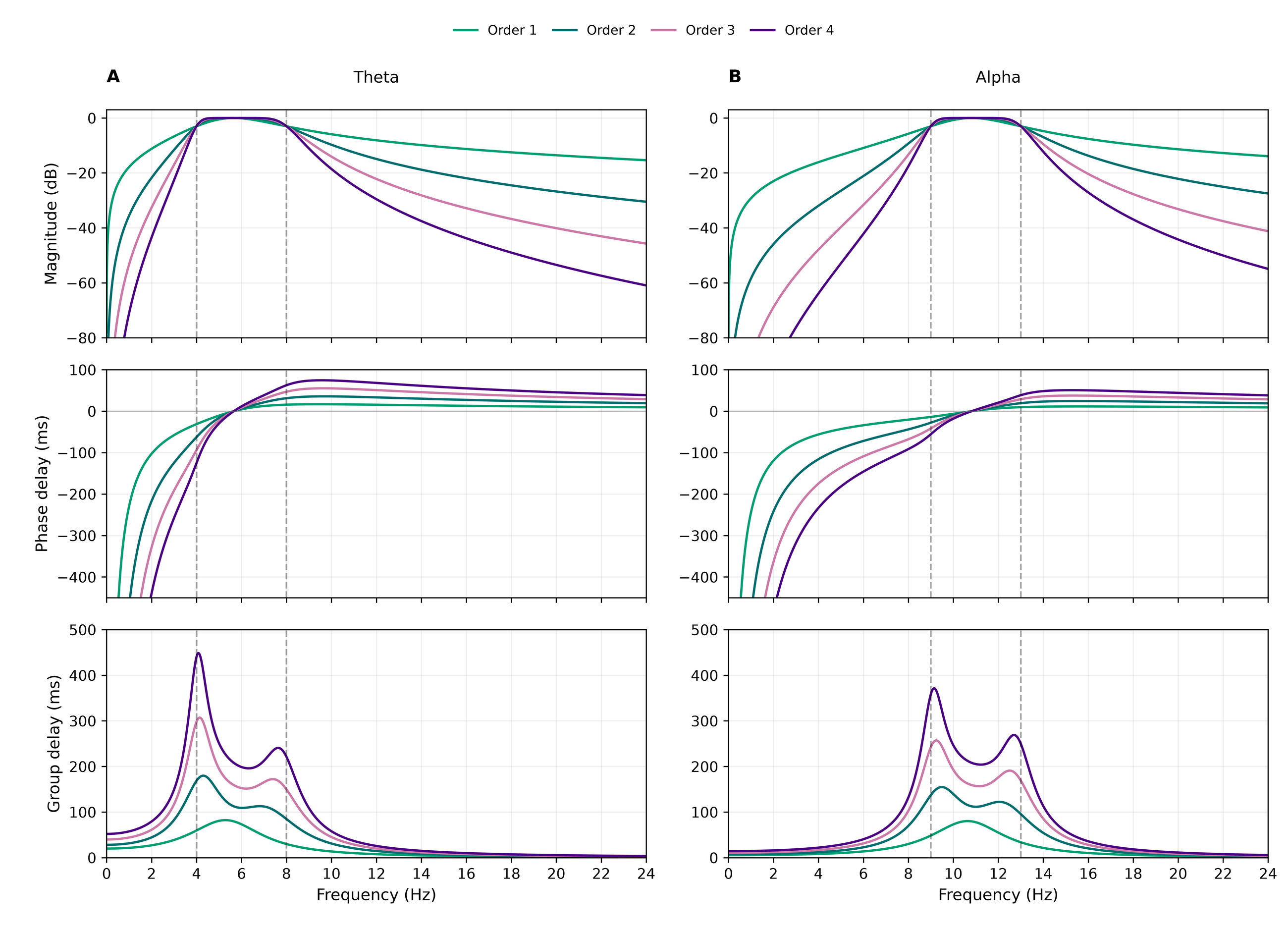}
    \caption{\textbf{(A)} Theta and \textbf{(B)} alpha bands forward pass Butterworth filter characteristic magnitude response, phase delay, and group delay.}
    \label{fig:panel_1}
\end{figure}

 We jointly optimize six per-subject parameters on the calibration split by minimizing calibration MACE ~\eqref{eq:optimization_objective}. We evaluated four optimizers: Gaussian-process Bayesian optimization using Skopt \texttt{gp\_minimize}~\citep{brochu2010tutorial}, Optuna \texttt{GPSampler}~\citep{akiba2019optuna}, Optuna \texttt{TPESampler}~\citep{bergstra2011algorithms} and random search~\citep{bergstra2012random}. Each optimizer was run with 20 exploratory and 100 objective evaluations across five random seeds per subject, with performance recorded at 10-evaluation intervals to characterize convergence rate, run-to-run variability, and asymptotic estimation accuracy under identical search spaces, data splits, and objective functions.

\begin{table}[H]
\centering
\caption{Optimized Parameters: Fixed Defaults and Per-Subject Search Ranges.}
\label{tab:ablation-params}
\begin{tabular}{lcccc}
\hline
\textbf{Parameter} & \textbf{Default($\theta$)} & \textbf{Search Range($\theta$)} & \textbf{Default($\alpha$)} & \textbf{Search Range($\alpha$)} \\
\hline
$p$          & $32$  & $[24, 100]$  & $96$   & $[48, 180]$   \\
$C$ (s)    & $1.0$ & $[0.8, 1.6]$ & $1.5$  & $[0.8, 2.0]$  \\
Filter Order       & $2$ & $[1, 4]$ & $2$  & $[1, 4]$ \\
Highpass (Hz)       & $4.0$ & $[4.0, 5.5]$ & $9.0$  & $[9.0, 10.5]$ \\
Lowpass (Hz)      & $8.0$ & $[6.0, 8.0]$ & $13.0$ & $[11.0, 13.0]$ \\
$P$ (samp)  & $60$  & $[30, 100]$  & $120$  & $[60, 180]$   \\
\hline
\end{tabular}
\end{table}

\subsection*{Comparators}

\subsubsection*{Phastimate}

Phastimate \citep{zrenner2020shaky} is the progenitor model developed
for the sensorimotor alpha rhythm. We use a Python port of the reference
MATLAB signal-processing implementation, with an integer genetic
algorithm for calibration. At each evaluation point, the real-time
window is demeaned and bandpass-filtered with a Hamming-windowed FIR
applied forward--backward (zero-phase). Edge samples are dropped from
each end to remove filter edge artifacts. A Yule--Walker AR model is
fit to the trimmed segment and forward-predicts
$\texttt{edge}+\lceil\texttt{hwin}/2\rceil$ samples. A Hilbert transform
of the final $\texttt{hwin}$ samples gives the instantaneous phase at
the sample corresponding to the evaluation point. Whereas the original
alpha pipeline individualizes the passband to the spectral peak
$\pm 1$~Hz, we fixed it to $4$--$8$~Hz for theta and $9$--$13$~Hz for
alpha. Following the reference calibration objective, the window length,
filter order, edge length and AR order were tuned per subject using a
Python genetic algorithm (population~100, integer genes, constraint
$\text{window}>3\times\text{filter order}$), minimizing circular
variance ($1-\text{PLV}$) over the upper-amplitude-half
calibration observations. Theta window-length, filter-order and
edge-length bounds were scaled from the original $1$~kHz settings to
the $300$~Hz sampling rate, while the AR-order range was retained.
The Hilbert window was $64$ samples for theta and $128$ samples for alpha.

\subsubsection*{PEAP}

A recent iteration of AR padding models introduces PEAP
\citep{kirchhoff2026phaseestimationautoregressivepadding}. Rather than
trimming the filtered signal before prediction, PEAP extends the raw
signal with Burg \cite{burg1975maximum} autoregression to forecast samples, shifting the filtering edge artifact onto the synthetic padding rather than the target sample. The padded signal is then bandpass-filtered forward--backward and Hilbert-transformed. A per-trial AR model generates the padding, avoiding the initial filtering and trimming used by Phastimate. We evaluated this model using grid-searched parameters (input length,
output length, AR order and AR model type) for both datasets, following
the released calibration procedure. 

\subsubsection*{EPN}

EPN \citep{liu2024epn} is a neural network originally proposed for
instantaneous phase estimation of the spatially filtered
\citep{mcfarland1997spatial,hjorth1975online} alpha rhythm
($8$--$13$~Hz). The architecture consists of a 1D average pooling layer
(kernel size~13), a fully connected hidden layer (500 units, ReLU
activation), dropout (rate 0.9), and a linear output layer producing
$(\sin\hat{\theta},\cos\hat{\theta})$. The input is $500$~ms of
backward-differenced raw EEG ending at the evaluation point. EPN was
reimplemented following the original implementation's hyperparameters:
Adam optimizer with learning rate $10^{-5}$, MSE loss, batch size~512,
and 3000 training epochs. Training samples were generated densely from
the calibration set at $10$~ms intervals across the $300$--$850$~ms
post-stimulus window for theta, and at $50$~ms intervals across
$1$--$4$~s for alpha, excluding the upper endpoint in each case. Input
features were z-scored using calibration-set statistics. The final
training weights were used for evaluation.

\subsubsection*{DBCNN}

DBCNN \citep{DBCNNHamdan} is a deep dual-branch complex neural network
developed for real-time instantaneous phase estimation. We adapted the
authors' released implementation (\href{https://github.com/Emadeldeen-Hamdan/DBCNN.git}{GitHub}), retaining its $512$-sample input and
network architecture. Each input contains only samples up to the
evaluation point, corresponding to approximately $1.71$~s for theta and
$512$~ms for alpha. The historical window is bandpass-filtered using
the released FFT-based squared-magnitude Butterworth response, adapted
to the sampling rate and target band. Two parallel branches each apply
two learned three-tap convolutions with downsampling. Their outputs are
concatenated and passed through a fully connected encoder, a
$16$-dimensional discrete cosine transform layer with learned weights
and thresholds, and a decoder producing the real and imaginary
components of the estimated analytic signal. Phase is obtained using
$\operatorname{atan2}$ of these components. Inputs were expressed in
microvolts and divided by $6453$, following the released implementation;
analytic-signal targets retained microvolt units. Models were trained
per subject using AdamW with learning rate $10^{-3}$, weight decay
$0.01$, batch size~512, and 7000 epochs, minimizing the summed MSE of
the real and imaginary components. Training windows used the same dense
calibration schedules as EPN. Twenty percent of calibration trials were
reserved for validation, and the weights with the lowest validation
loss were used for held-out evaluation.

\subsection*{Benchmarks}

To measure each model’s performance, the following metrics were applied: circular standard deviation (CSD), mean absolute circular error (MACE) \citep{mcintosh2020mlof, berens2009circstat, liu2024epn}, accuracy (ACC) \citep{shirinpour2020etp}, phase-locking value (PLV) \citep{berens2009circstat}, and absolute circular bias ($|\mathrm{Bias}|$).

\bigskip

\textbf{MACE:}
\begin{equation}
    \mathrm{MACE} = \frac{1}{K} \sum_{i=1}^{K} \left| \mathrm{Arg}\left( e^{j(\hat{\theta}_i - \theta_i)} \right) \right|
\end{equation}
where $\hat{\theta}_i$ and $\theta_i$ are the predicted and reference instantaneous phase, respectively. MACE ranges from 0\textdegree\ (perfect) to 180\textdegree\ (anti-phase).\\

\textbf{ACC:}
\begin{equation}
    \mathrm{ACC} = \left( 1 - \frac{\mathrm{MACE^\circ}}{180^\circ} \right) \times 100
\end{equation}

\textbf{PLV:}
\begin{equation}
    \mathrm{PLV} = \left| \frac{1}{K} \sum_{i=1}^{K} e^{j(\hat{\theta}_i - \theta_i)} \right|
\end{equation}

\textbf{CSD:}
\begin{equation}
    \mathrm{CSD} = \sqrt{-2 \ln (\mathrm{PLV})}
\end{equation}

CSD is calculated in radians and converted to degrees.\\

\textbf{$|\mathbf{Bias}|$:}
\begin{equation}
    |\mathrm{Bias}| =
    \left|
    \mathrm{Arg}\left(
        \frac{1}{K}\sum_{i=1}^{K}
        e^{j(\hat{\theta}_i-\theta_i)}
    \right)
    \right|.
\end{equation}\\

Absolute circular bias measures the magnitude of the systematic phase offset between predicted and reference phase, irrespective of its direction. It is calculated in radians and converted to degrees. Unlike MACE, which averages the magnitudes of individual phase errors, absolute circular bias is the magnitude of their circular mean direction. Consequently, a small absolute bias does not necessarily imply small individual errors.

All metrics are computed from held-out test predictions without post hoc bias correction. For each subject, errors are pooled across test trials and evaluation timepoints. Group results report the mean and sample standard deviation across subjects. For absolute circular bias, the absolute value is taken separately for each subject before group aggregation.

\newpage

\section{Results}

We hypothesized that the optimizer would improve upon our baseline in phase recovery, and capture the filter configuration that real-time systems prefer in signals with strong and inconsistent oscillatory response. In this section we present our initial findings from the ablation study and further decomposition of the Butterworth filter.

\subsection*{Optimization}

\begin{figure}[H]
    \centering
    \includegraphics[width=0.9\linewidth]{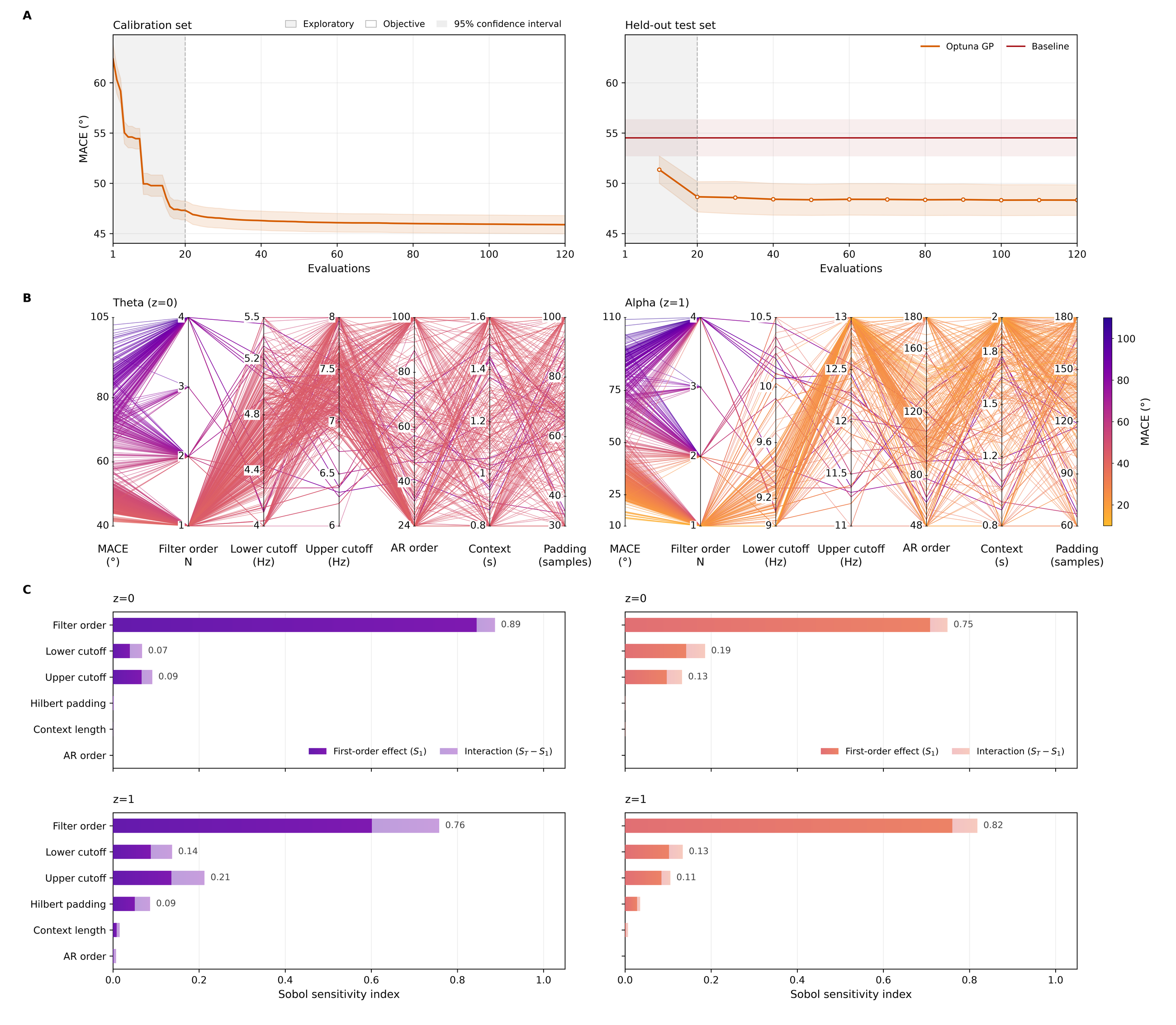}
    \caption{\textbf{Optimization characterization and parameter sensitivity of ablation study.} \textbf{(A)} Mean theta calibration MACE and corresponding held-out test MACE across \texttt{Optuna GPSampler} evaluations. Five optimizer seeds were averaged within each of 20 participants before group aggregation. \textbf{(B)} Parallel-coordinate plots of sampled parameter configurations for theta (left) and alpha (right), colored by calibration MACE. Each line represents one evaluated configuration. \textbf{(C)} Mean Sobol sensitivity indices across 20 participants for each band. Filter order was the dominant parameter in both bands and reconstruction modes. Here, $z=0$ denotes forward-only estimation and $z=1$ denotes backward reconstruction over the autoregressively predicted extension.}
    \label{fig:panel_2}
\end{figure}

Our ablation study shows that optimization reduced mean theta MACE
from $54.53^\circ$ to $48.33^\circ$ at 120 evaluations, averaging five
optimizer seeds within participants (Figure~\ref{fig:panel_2} A). Optimizer convergence occurs at roughly 20 objective evaluations. We deemed 20 exploratory and 20 objective evaluations to be suitable for the rest of the experiment. Mean reduction was $6.20^\circ$ (95\% CI $4.83$--$7.77^\circ$), with improvement in all 20 participants ($p<0.001$). Filter order had a larger total-order Sobol index than the strongest other parameter in every participant, in both bands and reconstruction modes (all $p<0.001$).

We compared forward-only ($z=0$) and backward-reconstruction
($z=1$) modes calibrated separately under equal 40-evaluation budgets.
Backward reconstruction reduced alpha MACE by $3.07^\circ$ (95\% CI $2.03$--$4.30^\circ$; Holm-adjusted $p<0.001$) but increased theta MACE by $1.81^\circ$
(95\% CI $0.84$--$2.78^\circ$; Holm-adjusted $p=0.00193$). Exact sign-test sensitivity checks supported both directions (alpha adjusted $p<0.001$; theta adjusted $p=0.00282$).

\subsection*{The Tradeoff of Non-Anticipative Filtering}

\begin{figure}[H]
    \centering
    \begin{minipage}[c]{.45\textwidth}
        \centering
        \includegraphics[
            width=\linewidth,
            trim={1cm 0.3cm 2cm 0.5cm},
            clip
        ]{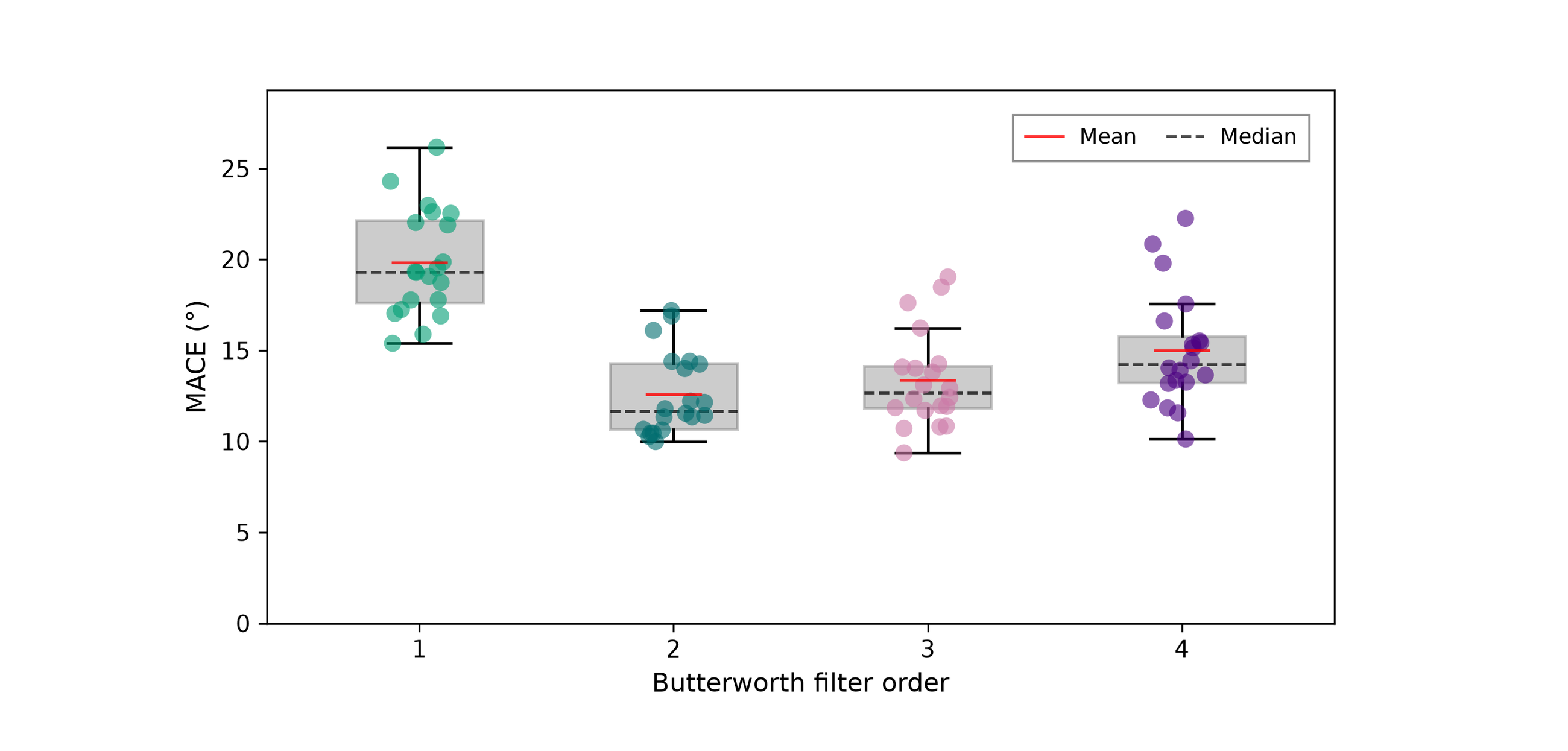}
    \end{minipage}%
    \hspace{.05\textwidth}%
   \begin{minipage}[c]{.30\textwidth}
    \captionsetup{justification=raggedright,singlelinecheck=false}
    \caption{\textbf{Offline theta-band phase error comparison} across Butterworth filter orders. Untuned 4--8 Hz filters were applied forward--backward to complete epochs, followed by the Hilbert transform. MACE was calculated relative to the FIR order-174 reference on held-out data from 20 subjects.}
    \label{fig:noncausal-order-comparison}
\end{minipage}
\end{figure}

With fixed $4$--$8$~Hz cutoffs and full-epoch forward--backward
filtering, theta mean MACE was $19.81^\circ$ for order 1,
$12.57^\circ$ for order 2, $13.37^\circ$ for order 3 and
$15.00^\circ$ for order 4
(Figure~\ref{fig:noncausal-order-comparison}).
Relative to order 1, orders 2--4 reduced mean MACE by
$7.24^\circ$ (adjusted CI $6.06$--$8.24^\circ$),
$6.44^\circ$ (adjusted CI $4.60$--$8.04^\circ$) and
$4.81^\circ$ (adjusted CI $2.72$--$6.61^\circ$), respectively
(all Holm-adjusted $p<0.001$). Exact sign-test checks also supported
all three advantages (all adjusted $p<0.001$).

\begin{figure}[H]
    \centering
    \includegraphics[width=0.9\linewidth]{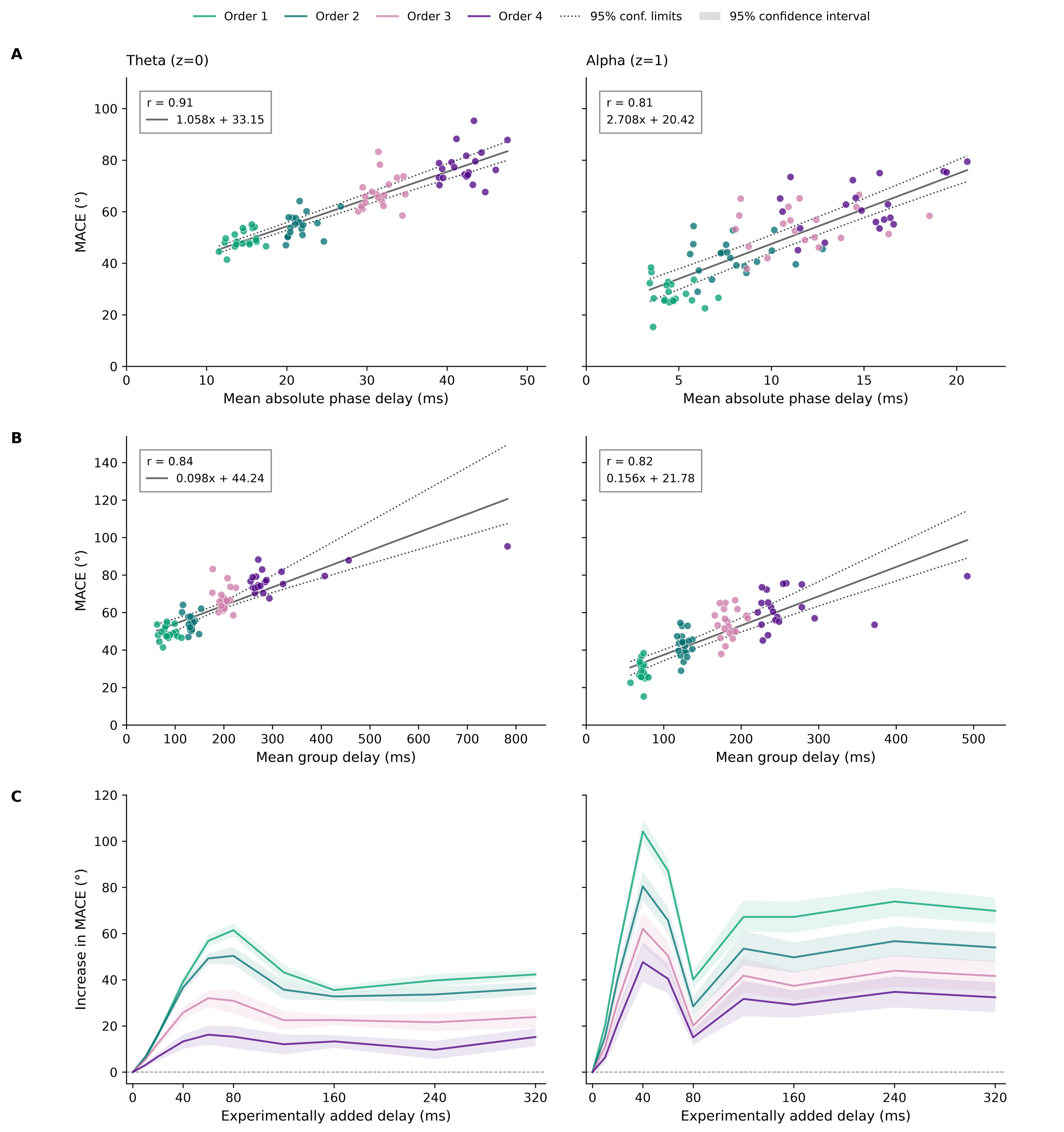}
    \caption{\textbf{Filter delay and real-time phase estimation error, with 20 participants per band}. \textbf{(A)} Held-out MACE versus mean absolute phase delay and \textbf{(B)} mean group delay of the forward Butterworth filter. \textbf{(C)} Mean paired change in MACE after inducing delay between forward filtering and AR estimation, relative to each participant's control delay baseline for the same filter order. Estimator parameters were held fixed. Theta uses forward-only estimation ($z=0$), while alpha retains backward reconstruction over the predicted extension ($z=1$).}
    \label{fig:panel_4}
\end{figure}

The reported order-search runs selected order 1 in all 20 participants
per band. In the held-out order comparisons, orders 2--4 increased mean
MACE relative to order 1 by $5.56$--$28.68^\circ$ in theta and
$14.68$--$34.48^\circ$ in alpha (all $p<0.001$;
Figure~\ref{fig:panel_4}). Within participants, MACE was positively
associated with forward-filter absolute phase delay (theta $r=0.976$;
alpha $r=0.973$) and group delay (theta $r=0.875$; alpha $r=0.927$;
all $p<0.001$).

For the frozen-estimator intervention, averaging the 10, 20 and 40~ms
conditions and four filter orders within participants, added delay
increased mean MACE by $15.61^\circ$ (95\% CI $14.37$--$16.90^\circ$)
in theta and $40.89^\circ$ (95\% CI $36.57$--$45.59^\circ$) in alpha.
Error was lower near a full cycle than near half a cycle by
$13.49^\circ$ (95\% CI $9.77$--$17.04^\circ$) in theta
(160 versus 80~ms) and $47.65^\circ$ (95\% CI
$39.43$--$56.39^\circ$) in alpha (80 versus 40~ms).
All four contrasts were significant ($p<0.001$).

\subsection*{Filter Reparameterization}

The consistent selection of order~1 motivated retaining a compact biquad and further characterizing its center frequency and decay time. An order-1 bandpass filter has a second-order transfer function and is therefore a biquad.

\begin{figure}[H]
    \centering
    \includegraphics[width=0.9\linewidth]{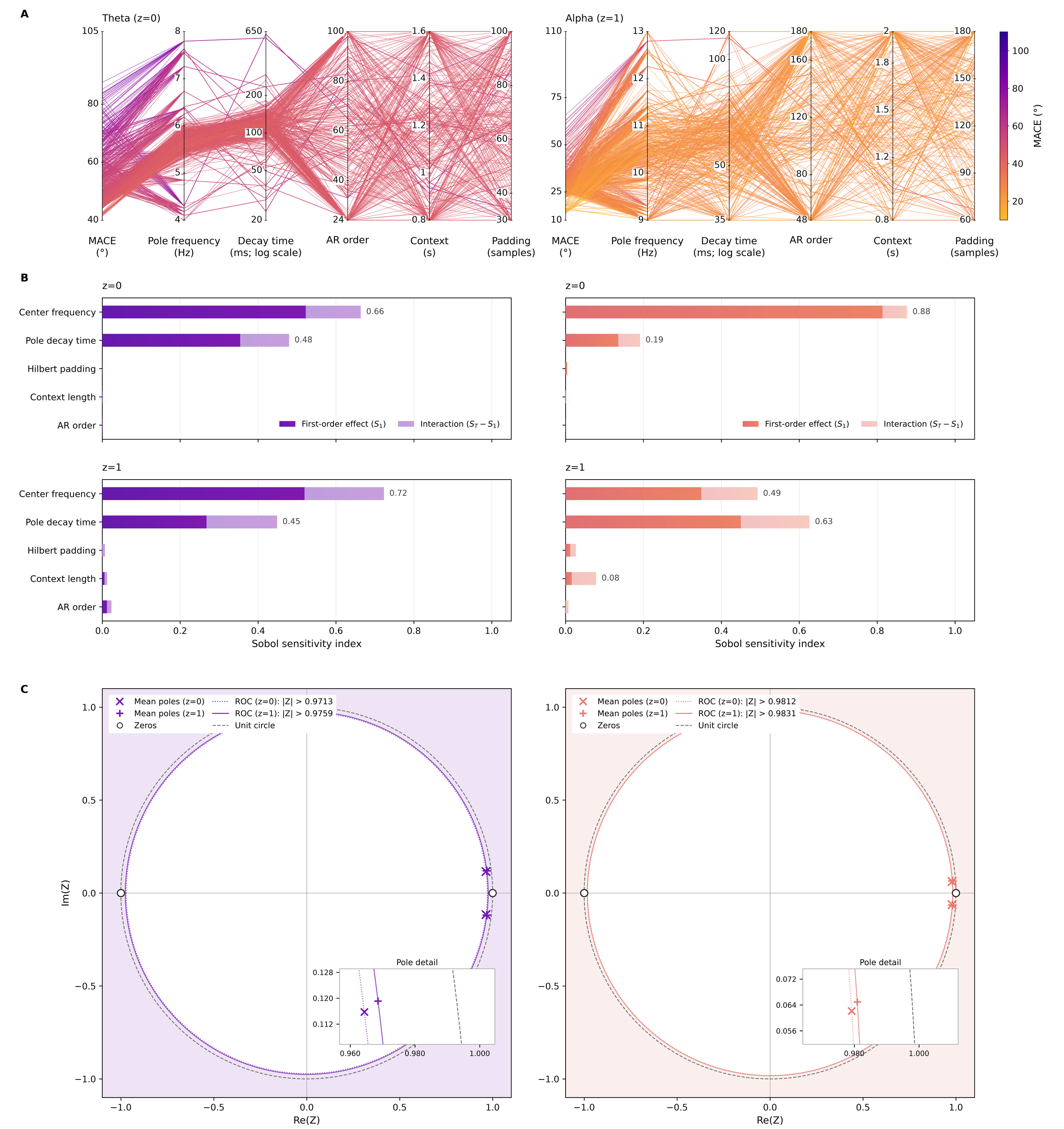}
    \caption{\textbf{Parameter sensitivity and pole geometry in BOB.} \textbf{(A)} Parallel-coordinate plots of sampled parameter configurations, colored by calibration MACE. Each line represents one evaluated configuration; decay time is displayed on a logarithmic scale. \textbf{(B)} Mean Sobol sensitivity indices across 20 participants per band. Center frequency and decay time contribute most strongly to calibration error. \textbf{(C)} Mean complex pole locations of the calibrated forward biquad sections for each reconstruction mode. Crosses denote $z=0$, plus signs denote $z=1$, and open circles show the fixed zeros at $\pm1$. The dashed gray circle marks the unit circle; colored boundaries and exterior shading indicate the regions of convergence associated with the mean pole pairs.}
    \label{fig:panel_5}
\end{figure}

Center frequency and decay time had the largest mean total-order Sobol
indices in the reparameterized biquad model; the mean padding-length
index remained at or below $0.027$ across both bands and reconstruction
modes (Figure~\ref{fig:panel_5}). Enabling backward reconstruction changed
decay-time importance relative to center-frequency importance by
$+0.819$ Sobol units in alpha (adjusted CI $0.670$--$0.936$) and
$-0.090$ in theta (adjusted CI $-0.148$--$-0.043$).

With reconstruction modes calibrated separately under the same
40-evaluation budget, backward reconstruction reduced alpha MACE by
$1.20^\circ$ (95\% CI $0.71$--$1.66^\circ$; $p=0.00145$) and
increased theta MACE by $1.96^\circ$ (95\% CI
$1.17$--$2.77^\circ$; $p<0.001$). For final evaluation we used the following reconstruction configurations: $z=0$ in the theta-band, and $z=1$ in the alpha band.

\subsection*{Performance}

\begin{figure}[H]
    \centering
    \includegraphics[width=0.8\linewidth]{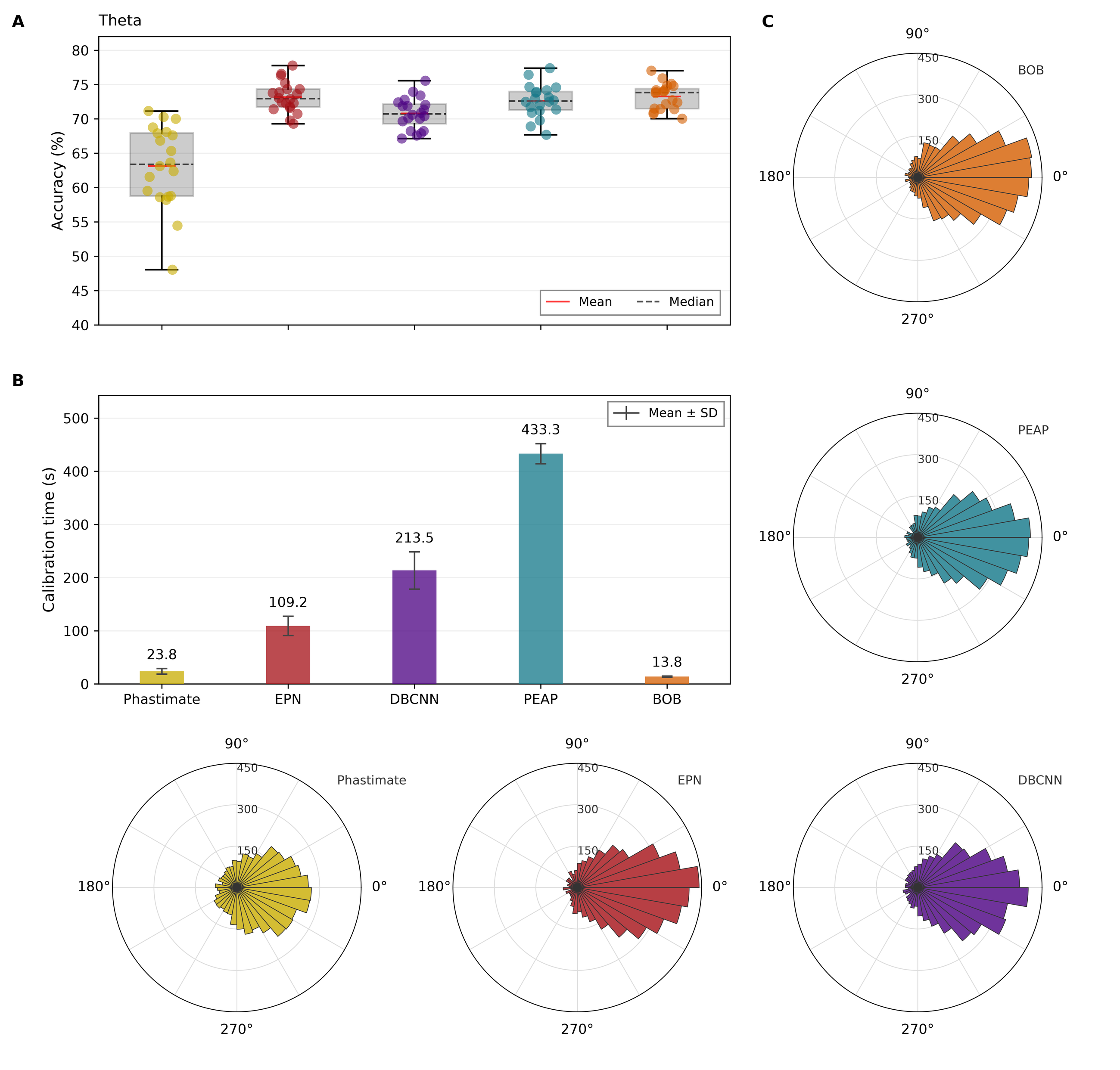}
    \caption{\textbf{Theta benchmark performance across 20 participants.} \textbf{(A)} Held-out phase accuracy. Points represent participants; boxes show the interquartile range, whiskers extend to observations within 1.5 interquartile ranges, red lines show means, and dashed lines show medians. \textbf{(B)} Mean calibration time with standard deviation. \textbf{(C)} Circular histograms of pooled held-out phase errors, using $10^\circ$ bins and a common count scale across methods. BOB uses forward-only estimation ($z=0$).}
    \label{fig:panel_6_theta}
\end{figure}

\begin{table}[H]
\centering
\caption{theta-band results. Values are mean $\pm$ sample standard deviation across 20 subjects. Calibration is done per-subject.}
\label{tab:theta_table}
\resizebox{\textwidth}{!}{%
\begin{tabular}{lcccccc}
\hline
\textbf{Model} &
\textbf{MACE ($^\circ$)} &
\textbf{ACC (\%)} &
\textbf{PLV} &
\textbf{CSD ($^\circ$)} &
\textbf{$|\mathrm{Bias}|$ ($^\circ$)} &
\textbf{Calibration (s)} \\
\hline
Phastimate & $66.34 \pm 10.79$ & $63.15 \pm 6.00$ & $0.35 \pm 0.12$ & $84.58 \pm 14.47$ & $22.28 \pm 25.04$ & $23.83 \pm 5.14$ \\
PEAP &
$49.29 \pm 4.24$ & $72.61 \pm 2.35$ & $0.55 \pm 0.05$ &
$62.64 \pm 4.69$ & $5.99 \pm 4.74$ & $433.34 \pm 18.88$ \\
DBCNN &
$52.57 \pm 4.09$ & $70.80 \pm 2.27$ & $0.51 \pm 0.05$ &
$66.71 \pm 4.95$ & $7.24 \pm 5.79$ & $213.55 \pm 35.01$ \\
EPN &
$48.31 \pm 3.98$ & $73.16 \pm 2.21$ & $0.56 \pm 0.05$ &
$61.66 \pm 4.65$ & $5.04 \pm 3.39$ & $109.19 \pm 18.11$ \\
\textbf{BOB} &
$\mathbf{48.14 \pm 3.39}$ & $\mathbf{73.25 \pm 1.88}$ &
$\mathbf{0.56 \pm 0.04}$ & $\mathbf{61.47 \pm 4.02}$ &
$\mathbf{4.50 \pm 3.24}$ & $\mathbf{13.77 \pm 1.13}$ \\
\hline
\end{tabular}%
}
\end{table}

\begin{figure}[H]
    \centering
    \includegraphics[width=0.8\linewidth]{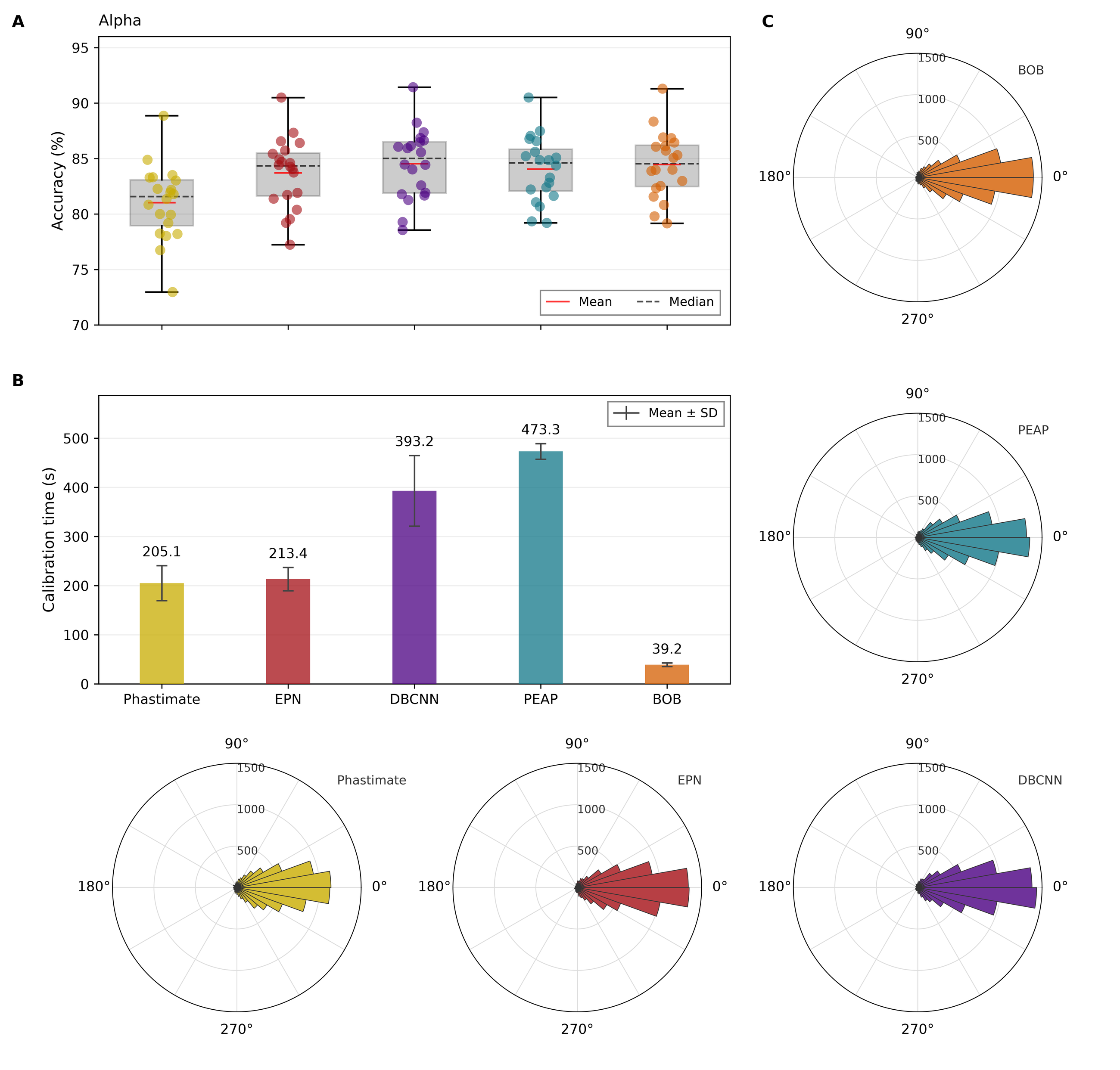}
    \caption{\textbf{Alpha benchmark performance across 20 participants.} \textbf{(A)} Held-out phase accuracy. Points represent participants; boxes show the interquartile range, whiskers extend to observations within 1.5 interquartile ranges, red lines show means, and dashed lines show medians. \textbf{(B)} Mean calibration time with standard deviation. \textbf{(C)} Circular histograms of pooled held-out phase errors, using $10^\circ$ bins and a common count scale across methods. BOB uses backward reconstruction ($z=1$).}
    \label{fig:panel_6_alpha}
\end{figure}

\begin{table}[H]
\centering
\caption{Alpha band results. Values are mean $\pm$ sample standard deviation across 20 subjects. Calibration is done per-subject.}
\label{tab:mu_table}
\resizebox{\textwidth}{!}{%
\begin{tabular}{lcccccc}
\hline
\textbf{Model} &
\textbf{MACE ($^\circ$)} &
\textbf{ACC (\%)} &
\textbf{PLV} &
\textbf{CSD ($^\circ$)} &
\textbf{$|\mathrm{Bias}|$ ($^\circ$)} &
\textbf{Calibration (s)} \\
\hline
Phastimate  & $34.15 \pm 6.05$ & $81.03 \pm 3.36$ & $0.73 \pm 0.07$ & $45.66 \pm 7.37$  & $2.51 \pm 1.56$ &  $205.11 \pm 35.65$ \\
PEAP &
$28.72 \pm 5.26$ & $84.05 \pm 2.92$ & $0.79 \pm 0.06$ &
$39.29 \pm 6.44$ & $2.50 \pm 1.60$ & $473.29 \pm 15.96$ \\
EPN &
$29.34 \pm 5.68$ & $83.70 \pm 3.15$ & $0.78 \pm 0.07$ &
$40.24 \pm 6.94$ & $2.74 \pm 2.10$ & $213.44 \pm 23.72$ \\
DBCNN &
$27.84 \pm 5.72$ & $84.53 \pm 3.18$ & $0.80 \pm 0.06$ &
$38.41 \pm 7.02$ & $2.41 \pm 1.51$ & $393.16 \pm 71.98$ \\
\textbf{BOB} &
$\mathbf{27.98 \pm 5.35}$ & $\mathbf{84.45 \pm 2.97}$ &
$\mathbf{0.79 \pm 0.06}$ & $\mathbf{38.57 \pm 6.67}$ &
$\mathbf{1.73 \pm 1.43}$ & $\mathbf{39.22 \pm 3.59}$ \\
\hline
\end{tabular}%
}
\end{table}

BOB achieved mean accuracy of $73.25\%$ in theta and $84.45\%$ in alpha,
ranking first and second, respectively, among the five evaluated
estimators (Figures~\ref{fig:panel_6_theta}--\ref{fig:panel_6_alpha}; Tables~\ref{tab:theta_table}--\ref{tab:mu_table}). In theta, mean accuracy exceeded
Phastimate by $10.11$ percentage points (adjusted CI $7.36$--$14.42$)
and DBCNN by $2.46$ points (adjusted CI $1.61$--$3.60$). In alpha,
the corresponding advantages were $3.43$ points over Phastimate
(adjusted CI $2.80$--$4.34$) and $0.76$ points over EPN
(adjusted CI $0.17$--$1.35$). Adjusted mean intervals included zero
for the remaining four comparisons. Across all eight comparisons,
the largest adjusted upper bound on mean BOB accuracy disadvantage
was $0.44$ percentage points, equivalent to $0.79^\circ$ MACE.

BOB had the shortest measured mean calibration time in both bands:
$13.77$~s in theta and $39.22$~s in alpha. Other methods required
$23.83$--$433.34$~s and $205.11$--$473.29$~s, respectively,
under the evaluated conditions (Tables~3--4).

Two-sided paired signed-rank comparisons, Holm-corrected across
all eight model comparisons, supported higher BOB accuracy than
Phastimate in both bands and DBCNN in theta (all $p<0.001$).
The alpha comparison against EPN was also significant ($p=0.024$).
The remaining comparisons were not significant: EPN in theta
and DBCNN in alpha (both $p=0.950$), and PEAP in both bands
(both $p=0.131$).

Sign-test sensitivity analyses corroborated the advantages over
Phastimate and theta DBCNN (all $p<0.001$), but not alpha EPN
($p=0.461$).

All five methods achieved higher mean accuracy in the evaluated alpha
cohort than in theta, with differences of $10.54$--$17.88$ percentage
points (all Holm-adjusted Welch $p<0.001$). BOB's difference was
$11.20$ percentage points (adjusted mean CI $9.24$--$13.21$).
The direction and significance persisted in the trimmed-mean
sensitivity analysis. These comparisons concern the two cohorts
and their respective recording conditions.

\begin{figure}[H]
    \centering
    \includegraphics[width=0.8\linewidth]{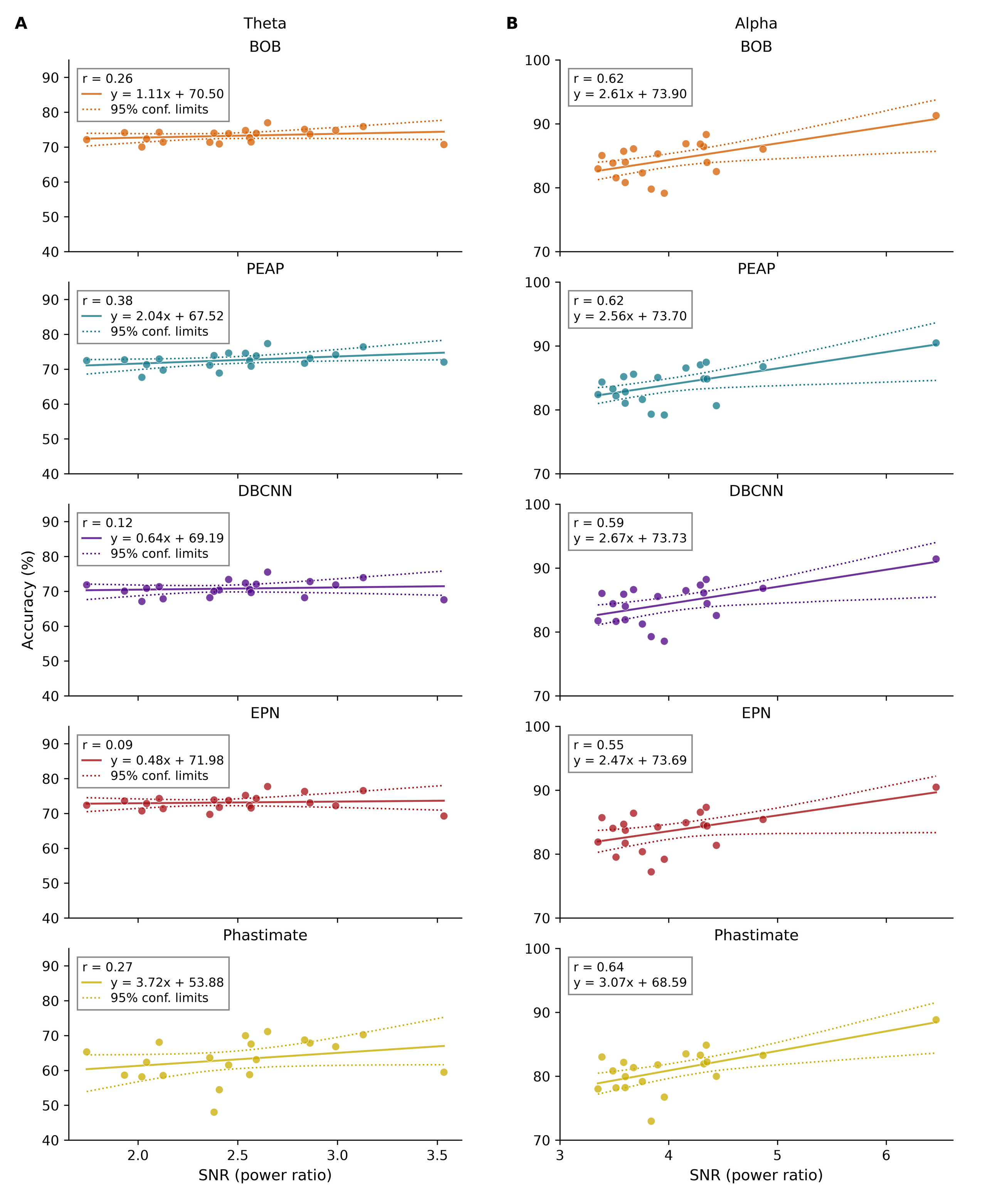}
    \caption{\textbf{SNR and held-out phase accuracy across the five estimators} in \textbf{(A)} theta and \textbf{(B)} alpha, with 20 participants per band.}
    \label{fig:panel_7}
\end{figure}

Across participants, BOB accuracy was positively associated with SNR
in alpha ($r=0.617$, 95\% CI $0.126$--$0.832$; adjusted $p=0.036$),
while the theta association was not significant
($r=0.258$, 95\% CI $-0.222$--$0.728$; adjusted $p=0.977$;
Figure~\ref{fig:panel_7}). Positive alpha associations also survived
correction for Phastimate, DBCNN and PEAP
(adjusted $p=0.030$--$0.038$); no theta association survived
correction. The direct between-cohort comparisons did not detect
differences in the Pearson correlations (all Fisher-$z$
Holm-adjusted $p=0.529$). Spearman sensitivity analyses did not
identify significant within-cohort associations after correction
(all adjusted $p\geq0.201$), and adjusted bootstrap intervals for
between-cohort Spearman differences included zero for every method.
The within-cohort associations between SNR and backward-reconstruction
benefit were also not significant (theta $r=0.013$, adjusted $p=0.956$; alpha $r=0.259$, adjusted $p=0.560$).

\subsubsection*{The Offline Reference}

\begin{figure}[H]
    \centering

    \begin{minipage}[c]{.6\textwidth}
        \begin{figure}[H]
        \centering
            \includegraphics[width=1\linewidth]{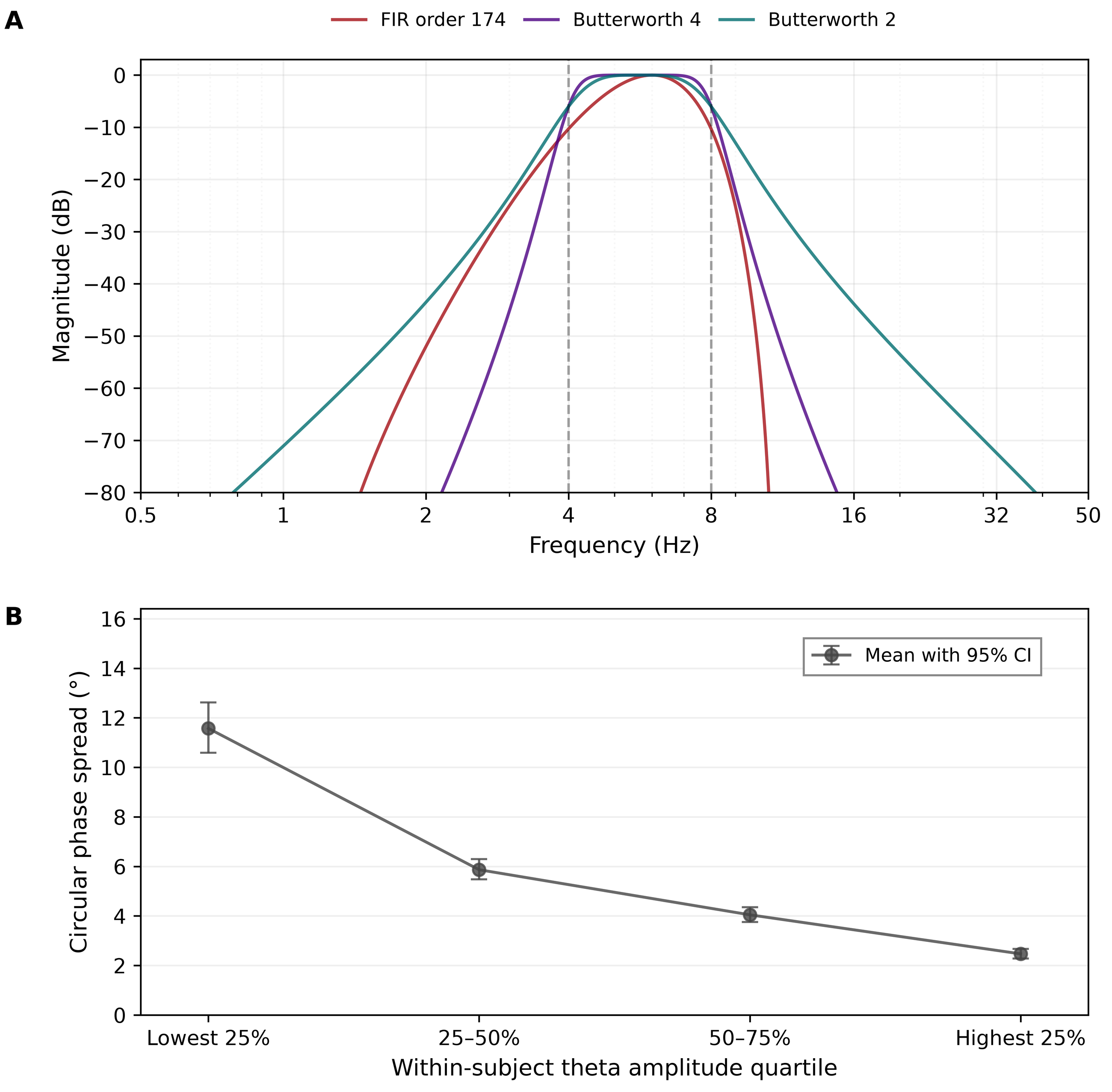}
            \label{fig:panel_8}
        \end{figure}
    \end{minipage}%
    \hspace{.05\textwidth}%
    \begin{minipage}[c]{.30\textwidth}
        \captionsetup{justification=raggedright,singlelinecheck=false}
        \caption{\textbf{Theta offline reference filter comparison.}
        \textbf{(a)} Zero-phase magnitude responses of the FIR order-174 and Butterworth order-2 and order-4 reference filters. \textbf{(b)} Circular phase spread across the three references to measure agreement; grouped by within-subject FIR amplitude quartile. Points show the mean of subject-level median spreads across 20 subjects; whiskers indicate 95\% subject-bootstrap confidence intervals.}
        \label{fig:gt-comparison}
    \end{minipage}

\end{figure}

After separate calibration against each reference, mean theta accuracy
was $73.25\%$ for the Hamming FIR reference, $72.97\%$ for second-order
Butterworth and $71.76\%$ for fourth-order Butterworth (Table \ref{tab:bob-theta-reference-sensitivity}).
The largest paired mean difference was $1.49$ percentage points
between FIR and fourth-order Butterworth (adjusted CI
$0.92$--$2.00$). Accuracy with the fourth-order reference was lower
than with either alternative (both $p<0.001$); the FIR versus
second-order comparison was not significant ($p=0.294$).

Mean participant-level median reference dispersion was $11.57^\circ$
in the lowest FIR-amplitude quartile and $2.47^\circ$ in the highest
(Figure~\ref{fig:gt-comparison}). The paired difference was $9.10^\circ$
(95\% CI $8.16$--$10.08^\circ$), with greater dispersion in the
lowest quartile in all 20 participants ($p<0.001$).

\begin{table}[H]
\centering
\caption{Theta results for BOB under three ground-truth definitions. Values are mean $\pm$ sample standard deviation across 20 subjects. All reference filters are applied forward--backward.}
\label{tab:bob-theta-reference-sensitivity}
\resizebox{\textwidth}{!}{%
\begin{tabular}{lllccccc}
\hline
\textbf{Reference filter} &
\textbf{Order} &
\textbf{MACE ($^\circ$)} &
\textbf{ACC (\%)} &
\textbf{PLV} &
\textbf{CSD ($^\circ$)} &
\textbf{$|\mathrm{Bias}|$ ($^\circ$)} \\
\hline
Hamming FIR & 174 &
$48.14 \pm 3.39$ & $73.25 \pm 1.88$ &
$0.56 \pm 0.04$ & $61.47 \pm 4.02$ &
$4.50 \pm 3.24$ \\
Butterworth & 2 &
$48.65 \pm 3.15$ & $72.97 \pm 1.75$ &
$0.56 \pm 0.04$ & $61.82 \pm 3.65$ &
$7.53 \pm 4.00$ \\
Butterworth & 4 &
$50.83 \pm 4.07$ & $71.76 \pm 2.26$ &
$0.53 \pm 0.05$ & $64.38 \pm 4.87$ &
$8.17 \pm 4.87$ \\
\hline
\end{tabular}%
}
\end{table}

\section{Discussion}

BOB achieved leading phase estimation performance across the evaluated theta and alpha scenarios, with substantially shorter calibration times than the other methods. Its mean accuracy ranked first in theta and second in alpha, where its MACE was only $0.14^\circ$ higher than DBCNN. The corrected comparisons supported advantages over Phastimate in both bands and DBCNN in theta, while the adjusted mean intervals also supported an alpha advantage over EPN (Tables~\ref{tab:theta_table}--\ref{tab:mu_table}).

In our analysis of filter behavior and the optimizer's search space, the findings point to several characterizations that are well motivated in the literature. Zrenner et al.\ \citep{zrenner2020shaky} describe the dependence of phase estimation accuracy on signal quality and the oscillatory activity present in the data. We observe a clear disparity between the spectra of our alpha and theta recordings, with a more distinct oscillatory peak in alpha (Figure~\ref{fig:supp2}). All five models performed significantly better in the evaluated alpha cohort, with mean accuracy advantages of $10.54$--$17.88$ percentage points (all adjusted $p<0.001$). The distinction between oscillatory power and the aperiodic background is relevant here. In visual working-memory recordings, van Engen et al.\ \citep{vanengen2026dissociating} found
detectable theta oscillations in $42\%$ of participants, compared with $91\%$ for alpha, after accounting for aperiodic activity. Donoghue et al.\ \citep{donoghue2022methodological} further show how weak oscillatory power and intermittent rhythms can complicate phase estimation. These findings provide a basis for considering the oscillatory structure of the signal when interpreting the difference between our two estimation settings.

The SNR results provide complementary evidence. BOB's alpha accuracy was positively associated with the reported SNR measure, as were the accuracies of Phastimate, DBCNN and PEAP after correction (Figure~\ref{fig:panel_7}). No theta association survived correction. This pattern is consistent with a clearer oscillatory response providing a more reliable basis for phase estimation. In stimulus-associated theta, a spectral summary over the analysed data may capture less of the local structure available at the estimation point. The correlation estimates were larger in alpha, although the direct between-cohort comparisons did not establish a difference. The observed pattern therefore motivates the interpretation of different estimation conditions without requiring theta accuracy to be independent of SNR.

In terms of configuration, the ablation study's sensitivity analysis identifies filter order as the dominant tuned parameter (Figure~\ref{fig:panel_2}). It exceeded the strongest other parameter in every participant, in both bands and reconstruction modes. The offline theta comparison helps explain why this is an important design choice. With fixed $4$--$8$~Hz cutoffs and access to complete epochs, orders 2--4, with greater spectral selectivity agreed more closely with the FIR reference than order 1 (Figure~\ref{fig:noncausal-order-comparison}). In real-time phase estimation all three higher orders increased error, and the optimizer selected order 1 throughout both cohorts (Figure~\ref{fig:panel_4}), indicating the proposed tradeoff. Thus, the filter response that best reproduces an offline target need not be the response best suited to real-time estimation.

Phase and group delay provide a temporal explanation for this tradeoff. Both were strongly associated with error across orders, and adding uncompensated delay to the frozen estimator increased error in both bands. We also observe lower error near one alpha cycle than near half a cycle (Figure~\ref{fig:panel_4}). A delayed waveform can return to a similar circular phase after one cycle, but the estimate still refers to an earlier signal state. For an evolving response, that distinction matters. These results support treating temporal alignment as a central consideration alongside spectral selectivity \cite{wodeyar2023different, zrenner2020shaky} when choosing
a filter for real-time phase estimation.

We further characterized this tradeoff by decomposing the biquad filter into center frequency and decay time. These parameters describe two useful filter properties: where the response is concentrated within the band, and how strongly past samples continue to influence the current output. In alpha, enabling backward reconstruction shifted sensitivity strongly toward decay time relative to center frequency (Figure~\ref{fig:panel_5}). This points to the importance of the filter's memory in the combined estimator. Decay time also changes bandwidth, so its temporal and spectral effects are considered together. BOB can adjust these properties within a compact second-order filter instead of relying on an increase in order to shape its response.

The reconstruction results fit this interpretation. Backward filtering improved alpha accuracy but reduced theta accuracy under separate calibration with equal budgets. We interpret this as a case where a
more sustained oscillatory signal allows the past to provide a more useful guide to the future. Burg continuation supplies the predicted samples used by the backward pass. When that continuation represents the upcoming waveform well, it can support useful backward correction; when the local response changes, the same operation can propagate continuation errors into the phase estimate. The spectral findings and the work of van Engen et al.\ and Donoghue et al.\ \citep{vanengen2026dissociating,donoghue2022methodological} support oscillatory structure as an explanation for this contrast. SNR describes power in the target band relative to background activity, while the usefulness of the continuation also depends on its temporal consistency. Oscillatory power and continuity are therefore relevant considerations when choosing reconstruction mode. Our findings suggest that in practice, reconstruction mode should be chosen for the intended signal and task. The low sensitivity to padding length over the tested positive ranges also places the emphasis on the behavior of the backward reconstruction and filter, rather than on padding duration alone.

Furthermore, BOB's adaptive nature allowed it to perform well with different offline references after recalibration (Table~\ref{tab:bob-theta-reference-sensitivity}). Mean theta accuracy remained between $71.76\%$ and $73.25\%$ across the three definitions. Although the fourth-order Butterworth target produced lower accuracy, the observed $1.49$-percentage-point range supports BOB's ability to adapt across the tested references. The references themselves also disagreed more at low amplitude (Figure~\ref{fig:gt-comparison}), consistent with the uncertainty in offline phase targets described by Zrenner
et al.\ \citep{zrenner2020shaky}. This makes reference selection part of the estimation problem and motivates calibrating the model to the target appropriate for the application.

We think BOB's calibration time and modular approach make a clear case for evaluating deployment on small-form-factor hardware. Mean calibration times were $13.77$~s in theta and $39.22$~s in alpha. The compact biquad filter and Burg prediction, together with the separation of optimization from subsequent estimation, allow calibration and deployment to be organized as separate computational tasks.

\subsection*{Limitations and Future Directions}

BOB has not yet been validated in a true online closed-loop system. The immediate deployment work is to evaluate continuous tracking, filter-state handling and phase error at actual stimulation delivery. Inference latency, timing variability and power consumption can all be quantified on the intended hardware. Calibration also requires prerecorded participant-specific data; the reported calibration times measure computation, rather than the recording duration required to obtain those data.

The two cohorts represent different recording conditions. In particular, the alpha participants were selected for high SNR, whereas theta was evaluated at five post-stimulus time points on an individually selected channel. The between-cohort accuracy difference therefore describes the
evaluated scenarios. The proposed connection between oscillatory structure, continuation quality and backward reconstruction draws on our measured filter and reconstruction effects and the cited literature.

The reference phase is defined by an offline signal-processing procedure, and its biological significance is task dependent and not established. Agreement with that reference provides a benchmark for estimation performance. Establishing useful neurofeedback or stimulation
requires testing whether tracking the selected phase produces a reliable physiological or behavioral effect. Our immediate future direction is BOB's deployment in a modular small-form-factor closed-loop system to continuously estimate theta phase during stimulus-associated activity. The theta-controlled neurofeedback work of Demarest et al.\ \citep{demarest2024theta} motivates pain-related applications. Extending this approach from theta-power feedback to phase-contingent stimulation would allow us to test whether the additional timing information contributes to neural modulation or pain outcomes.

\subsection*{Conclusion}

BOB combines a compact biquad--filter Burg autoregressive estimator with Bayesian optimization to achieve leading accuracy among the evaluated methods and the shortest measured mean calibration time in both datasets. Corrected comparisons establish advantages over specific comparators, while adjusted intervals quantify the remaining uncertainty among the leading methods. The filter and reconstruction analyses support calibrating temporal and spectral properties jointly, with the backward reconstruction mode selected for the intended task.

\subsection*{Author Contributions}

\begin{enumerate}
    \item Ivan Martinez-Kay:
    \begin{itemize}
        \item Conceptualization, Methodology, Software, Validation, Visualization, and Formal analysis
        \item Writing - Original draft, Project administration, and Editing
    \end{itemize}
    \item Tatiana Kardashina:
    \begin{itemize}
        \item Data curation and Pre-processing
        \item Writing - Review and Editing
    \end{itemize}
    \item ShiNung Ching, PhD:
    \begin{itemize}
        \item Supervision and Visualization
        \item Writing - Review and Editing
    \end{itemize}
    \item Eric C. Leuthardt, MD, MBA:
    \begin{itemize}
        \item Supervision and Funding acquisition
        \item Writing - Review and Editing
    \end{itemize}
    \item Benjamin T. Acland, PhD:
    \begin{itemize}
        \item  Supervision, Funding acquisition, Visualization, Data curation and Pre-processing
        \item Writing -  Review and Editing
    \end{itemize}
\end{enumerate}

\subsection*{Hardware Used}

Model training and calibration were performed on a single local workstation with an AMD Ryzen AI 7 350 CPU (8C/16T), 16GB DDR5 RAM, and an NVIDIA RTX 5060 Laptop GPU (8GB GDDR7), without reliance on cloud compute or distributed training infrastructure.

\subsection*{Code and Data Availability}
The codebase used in this study is available on \href{https://github.com/IvanMK518/BOB.git}{GitHub}. Access to stimulus-locked raw data may be granted upon reasonable request to the corresponding author \href{mailto:leuthardte@wustl.edu}{leuthardte@wustl.edu}.

\subsection*{Acknowledgements}

This work was supported by The Taylor Family Department of Neurosurgery's division of neurotechnology, and The Preston M. Green Department of Electrical and Systems Engineering. The methods described in this work are the subject of \texttt{U.S. Provisional Patent Application No.~64/187,116 filed October 29, 2026}.

We would like to thank the late Dr. Alessandro Veneziani of Emory University for instructing the foundational mathematics and discussions regarding band-pass filtering, time domain analysis, and Fourier transform-based techniques. At the time of writing, we have regretfully been informed of his unexpected passing. We remember him with gratitude for his teaching, guidance, and lasting influence on this work.

\section*{Supplementary Figures}

\begin{figure}[H]
    \centering
    \includegraphics[width=1\linewidth]{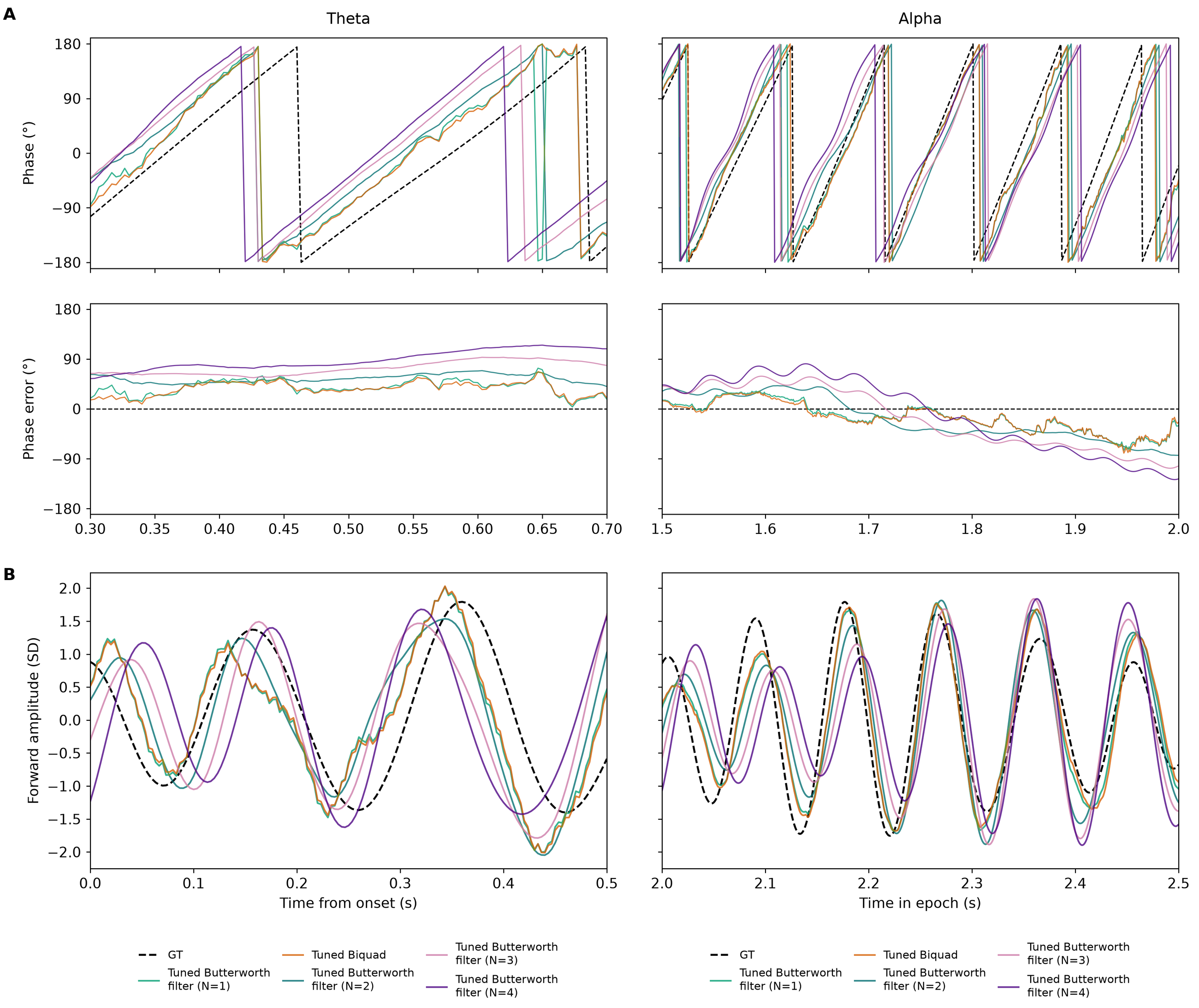}
    \caption{\textbf{Time domain phase and waveform comparisons from held-out trials.} \textbf{(A)} Instantaneous phase estimates (upper panels) and signed circular phase errors relative to the offline reference (lower panels). Estimates use our final benchmarked model configurations. \textbf{(B)} Forward-filtered waveforms, with each trace standardized to zero mean and unit standard deviation over the displayed interval. N corresponds to filter order.}
    \label{fig:supp1}
\end{figure}

\begin{figure}[H]
    \centering
    \includegraphics[width=1\linewidth]{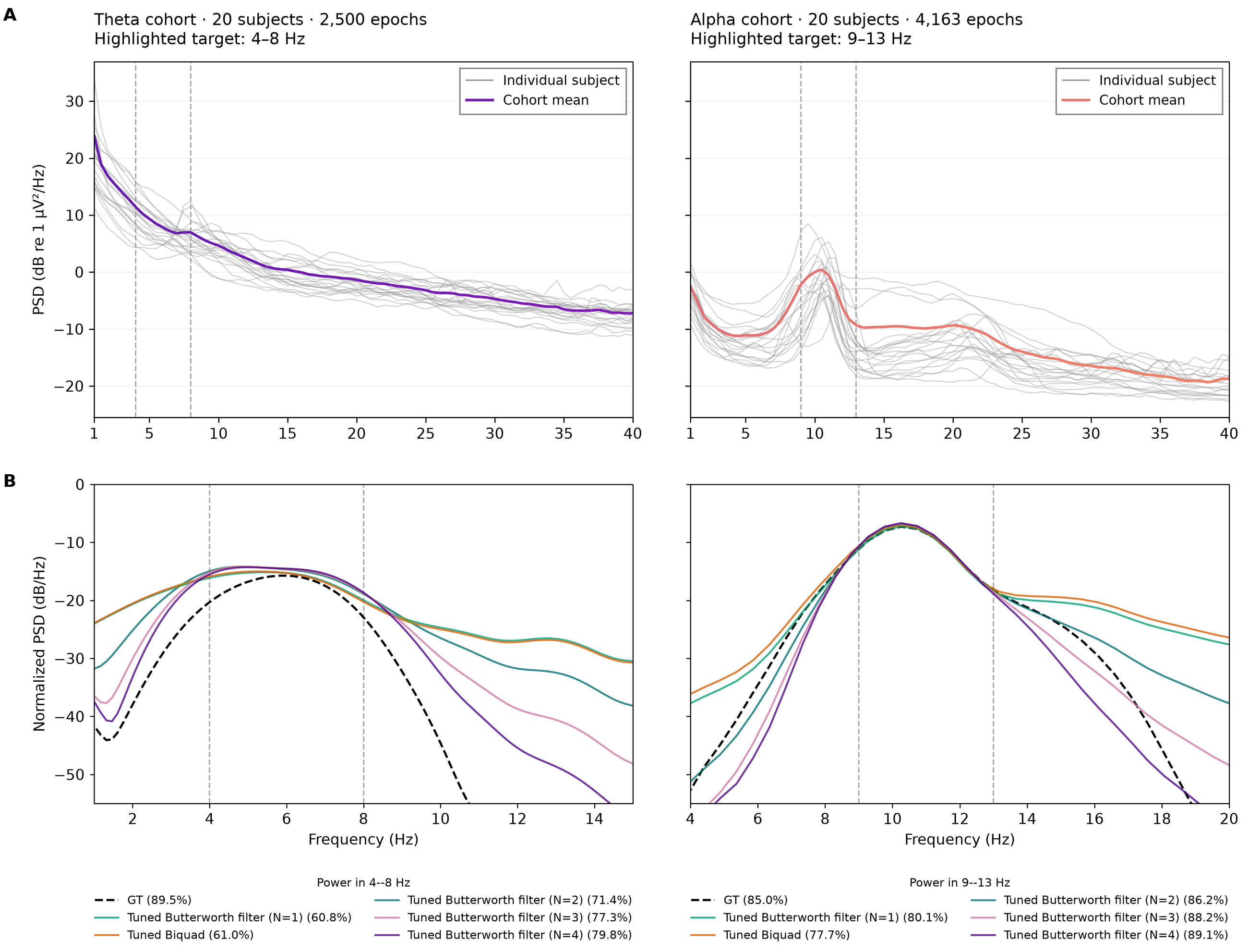}
    \caption{\textbf{Power Spectra Characteristics.} \textbf{(A)} Broadband power spectral densities across 20 participants per band. Spectra include calibration and test epochs and were estimated using 2~s Welch windows with 50\% overlap. \textbf{(B)} Normalized spectra for the example participants in Figure \ref{fig:supp1}, averaged over matched observed windows across held-out trials and evaluation points. Spectra share a normalizer based on the input signal's 1--40~Hz power. Legend percentages indicate the proportion of each output's 1--40~Hz power within the target band.}
    \label{fig:supp2}
\end{figure}

\begin{figure}[H]
    \centering
    \includegraphics[width=0.7\linewidth]{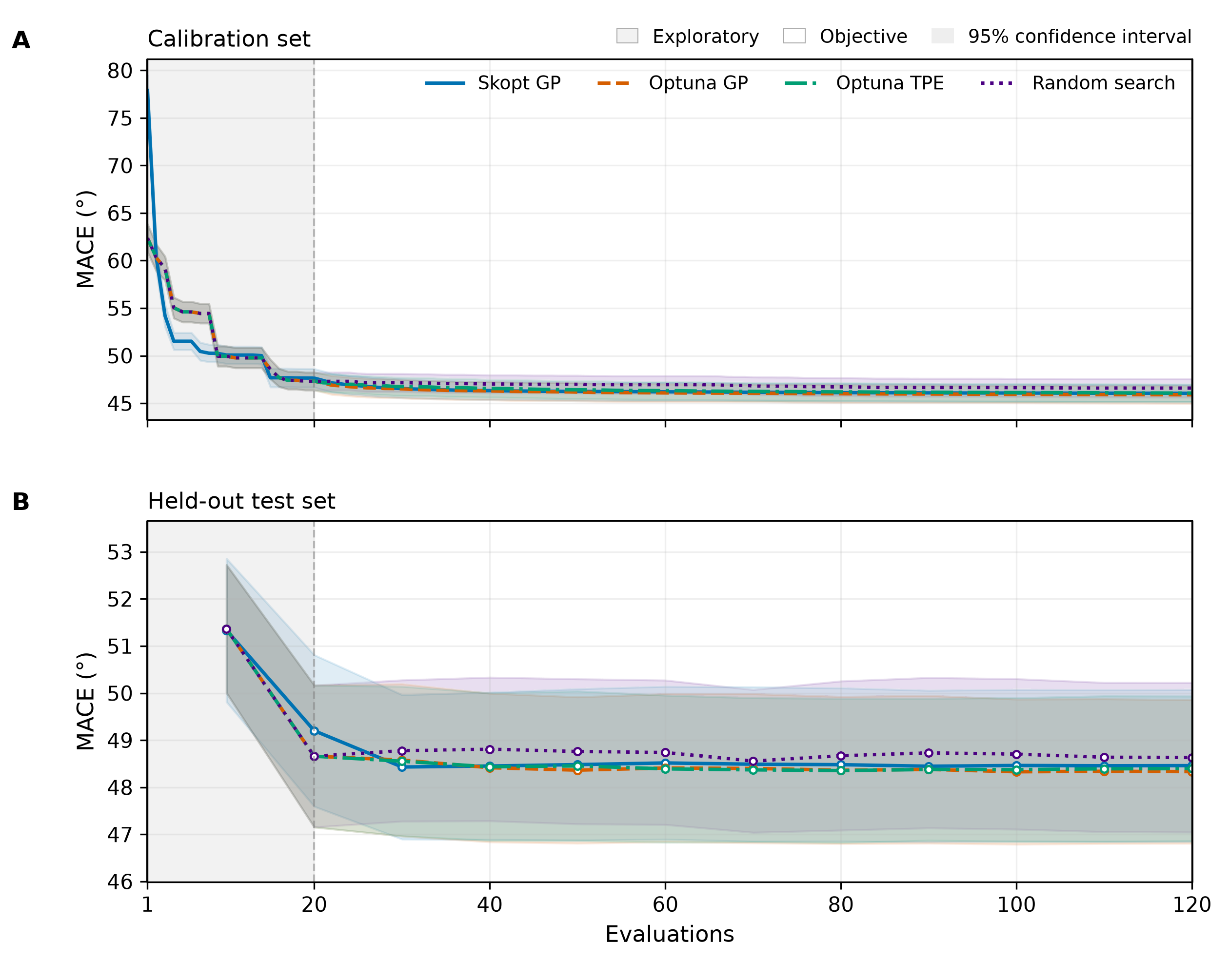}
    \caption{\textbf{Optimizer convergence in ablation study for theta phase estimation.} \texttt{Skopt GP, Optuna GP, Optuna TPE} and \texttt{random search} were evaluated under identical search spaces and data splits. \textbf{(A)} Best calibration MACE obtained through each evaluation. \textbf{(B)} Held-out test MACE of the calibration-selected configuration, evaluated at 10-evaluation intervals. Seeds were averaged within each of 20 participants before group aggregation.}
    \label{fig:supp3}
\end{figure}

\begin{table}[H]
\centering
\caption{BOB optimized per-session held-out test parameters per subject ($N$). AR Order ($p$), context duration ($C$), center frequency ($f_c$), decay time constant ($\tau$), and Hilbert padding in samples ($P$).}
\label{tab:bob_0ms}
\hspace*{-1.2cm}%
\begin{subtable}[t]{0.4\textwidth}
    \caption{Theta (MACE: 48.14\textdegree, PLV: 0.563)}
    \label{tab:bob_theta_params}
    \footnotesize
    \begin{tabular}{cccccc}
    \hline
    \textbf{$N$} & \textbf{$p$} & \textbf{$C$ (s)} & \textbf{$f_c$ (Hz)} & \textbf{$\tau$ (ms)} & \textbf{$P$ (samples)} \\
    \hline
    1  &  70 & 1.1 & 5.6 & 122.8 &  65 \\
    2  &  73 & 1.1 & 5.8 & 118.1 &  90 \\
    3  & 100 & 1.2 & 5.9 & 143.1 &  64 \\
    4  & 100 & 1.6 & 5.4 & 127.0 & 100 \\
    5  &  68 & 0.8 & 5.9 & 155.0 &  30 \\
    6  &  46 & 1.2 & 5.6 & 105.0 &  70 \\
    7  &  24 & 1.4 & 5.6 &  92.4 &  30 \\
    8  &  24 & 0.8 & 5.5 &  88.9 & 100 \\
    9  &  39 & 1.6 & 5.8 & 118.9 &  69 \\
    10 &  80 & 0.8 & 5.8 & 149.0 &  48 \\
    11 &  55 & 1.3 & 5.8 & 125.6 &  60 \\
    12 &  24 & 1.6 & 5.8 & 125.1 &  30 \\
    13 &  24 & 0.8 & 5.9 &  90.1 &  80 \\
    14 &  38 & 0.9 & 5.4 & 127.3 &  30 \\
    15 &  51 & 1.4 & 5.8 & 121.0 &  80 \\
    16 &  24 & 1.1 & 5.8 & 115.4 &  81 \\
    17 &  24 & 1.6 & 5.4 & 134.2 & 100 \\
    18 &  85 & 1.2 & 6.0 &  81.0 &  57 \\
    19 &  41 & 1.1 & 5.6 &  94.2 & 100 \\
    20 &  24 & 0.8 & 5.6 & 125.3 &  32 \\
    \hline
    \end{tabular}
\end{subtable}
\hspace{45pt}
\begin{subtable}[t]{0.4\textwidth}
    \caption{Alpha (MACE: 27.98\textdegree, PLV: 0.794)}
    \label{tab:bob_mu_params}
    \footnotesize
    \begin{tabular}{cccccc}
    \hline
    \textbf{$N$} & \textbf{$p$} & \textbf{$C$ (s)} & \textbf{$f_c$ (Hz)} & \textbf{$\tau$ (ms)} & \textbf{$P$ (samples)} \\
    \hline
    1  & 104 & 2.0 & 10.9 & 62.3 & 165 \\
    2  & 180 & 2.0 & 10.6 & 47.5 & 180 \\
    3  & 180 & 2.0 & 10.8 & 64.2 & 165 \\
    4  & 107 & 1.5 & 10.8 & 47.5 & 153 \\
    5  & 109 & 1.6 & 10.3 & 68.3 & 180 \\
    6  &  99 & 1.1 & 10.4 & 49.5 & 160 \\
    7  &  95 & 1.7 & 10.8 & 64.4 & 180 \\
    8  & 116 & 2.0 & 11.0 & 69.9 & 135 \\
    9  & 113 & 1.3 & 10.6 & 62.1 & 140 \\
    10 & 168 & 2.0 & 10.6 & 64.3 & 180 \\
    11 & 180 & 1.6 &  9.7 & 69.8 & 180 \\
    12 & 129 & 1.0 & 10.5 & 61.4 &  95 \\
    13 & 180 & 2.0 & 10.5 & 79.0 & 180 \\
    14 & 107 & 1.3 & 10.3 & 53.7 & 112 \\
    15 &  50 & 2.0 &  9.8 & 43.8 & 160 \\
    16 &  99 & 1.1 & 10.4 & 49.5 & 160 \\
    17 & 142 & 1.5 & 10.7 & 61.9 & 122 \\
    18 &  99 & 1.1 & 10.4 & 49.5 & 160 \\
    19 & 146 & 2.0 & 10.9 & 73.5 & 150 \\
    20 &  97 & 1.3 & 10.2 & 60.2 & 105 \\
    \hline
    \end{tabular}
\end{subtable}
\end{table}

\bibliographystyle{unsrtnat}
\bibliography{references}

@article{liu2024epn,
  author = {Liu, Hao and Qi, Zihui and others},
  title = {A Novel Real-time Phase Prediction Network in {EEG} Rhythm},
  journal = {Neuroscience Bulletin},
  volume = {41},
  number = {3},
  pages = {391--405},
  year = {2024},
  doi = {10.1007/s12264-024-01321-z}
}

@article{shirinpour2020etp,
  author = {Shirinpour, Sina and Alekseichuk, Ivan and Mantell, Kathleen and Opitz, Alexander},
  title = {Experimental Evaluation of Methods for Real-Time {EEG} Phase-Specific Transcranial Magnetic Stimulation},
  journal = {Journal of Neural Engineering},
  volume = {17},
  number = {4},
  pages = {046002},
  year = {2020},
  doi = {10.1088/1741-2552/ab9dba}
}

@article{shirinpour2025btp,
  author = {Shirinpour, Sina and Alekseichuk, Ivan and G{\"u}th, Malte R. and Haigh, Zachary J. and Wischnewski, Miles and Opitz, Alexander},
  title = {Bayesian Temporal Prediction: {A} Robust Algorithm for Real-time {EEG} Phase-dependent Brain Stimulation},
  journal = {IEEE Transactions on Biomedical Engineering},
  volume = {73},
  number = {2},
  pages = {611--620},
  year = {2025},
  doi = {10.1109/TBME.2025.3589970}
}

@article{kim2023eeg,
  author = {Kim, Brian and Erickson, Brian A. and others},
  title = {{EEG} Phase Can Be Predicted with Similar Accuracy across Cognitive States after Accounting for Power and Signal-to-Noise Ratio},
  journal = {eNeuro},
  volume = {10},
  number = {9},
  year = {2023},
  doi = {10.1523/ENEURO.0050-23.2023}
}

@article{zrenner2020shaky,
  author = {Zrenner, Christoph and Galevska, Darinka and Nieminen, Jaakko O. and Baur, David and Stefanou, Maria-Ioanna and Ziemann, Ulf},
  title = {The Shaky Ground Truth of Real-Time Phase Estimation},
  journal = {NeuroImage},
  volume = {214},
  pages = {116761},
  year = {2020},
  doi = {10.1016/j.neuroimage.2020.116761}
}

@article{zrenner2018realtime,
  author = {Zrenner, Christoph and Desideri, Debora and Belardinelli, Paolo and Ziemann, Ulf},
  title = {Real-time {EEG}-Defined Excitability States Determine Efficacy of {TMS}-Induced Plasticity in Human Motor Cortex},
  journal = {Brain Stimulation},
  volume = {11},
  number = {2},
  pages = {374--389},
  year = {2018},
  doi = {10.1016/j.brs.2017.11.016}
}

@article{mcintosh2020mlof,
  author = {McIntosh, James R. and Sajda, Paul},
  title = {Estimation of Phase in {EEG} Rhythms for Real-time Applications},
  journal = {Journal of Neural Engineering},
  volume = {17},
  number = {3},
  pages = {034002},
  year = {2020},
  doi = {10.1088/1741-2552/ab8683}
}

@article{chen2013realtime,
  author = {Chen, Liang L. and Madhavan, Radhika and Rapoport, Benjamin I. and Anderson, William S.},
  title = {Real-time Brain Oscillation Detection and Phase-locked Stimulation Using Autoregressive Spectral Estimation and Time-Series Forward Prediction},
  journal = {IEEE Transactions on Biomedical Engineering},
  volume = {60},
  number = {3},
  pages = {753--762},
  year = {2013},
  doi = {10.1109/TBME.2012.2226595}
}

@article{berens2009circstat,
  author = {Berens, Philipp},
  title = {{CircStat}: {A} {MATLAB} Toolbox for Circular Statistics},
  journal = {Journal of Statistical Software},
  volume = {31},
  number = {10},
  pages = {1--21},
  year = {2009},
  doi = {10.18637/jss.v031.i10}
}

@article{donoghue2020fooof,
  author = {Donoghue, Thomas and Haller, Matar and Peterson, Erik J. and Varma, Paroma and Sebastian, Priyadarshini and Gao, Richard and Noto, Torben and Lara, Antonio H. and Wallis, Joni D. and Knight, Robert T. and Shestyuk, Avgusta and Voytek, Bradley},
  title = {Parameterizing Neural Power Spectra into Periodic and Aperiodic Components},
  journal = {Nature Neuroscience},
  volume = {23},
  number = {12},
  pages = {1655--1665},
  year = {2020},
  doi = {10.1038/s41593-020-00744-x}
}

@article{demarest2024theta,
  author = {Demarest, Phillip and Rustamov, Nabi and Swift, James and Xie, Tao and Adamek, Markus and Cho, Hohyun and Wilson, Elizabeth and Han, Zhuangyu and Belsten, Alexander and Luczak, Nicholas and Brunner, Peter and Haroutounian, Simon and Leuthardt, Eric C.},
  title = {A Novel Theta-Controlled Vibrotactile Brain--Computer Interface to Treat Chronic Pain: {A} Pilot Study},
  journal = {Scientific Reports},
  volume = {14},
  number = {1},
  pages = {3433},
  year = {2024},
  doi = {10.1038/s41598-024-53261-3}
}

@phdthesis{burg1975maximum,
  author = {Burg, John Parker},
  title = {Maximum Entropy Spectral Analysis},
  school = {Stanford University},
  address = {Stanford, CA},
  year = {1975}
}

@book{kay1988modern,
  author = {Kay, Steven M.},
  title = {Modern Spectral Estimation: Theory and Application},
  year = {1988},
  publisher = {Prentice Hall},
  address = {Englewood Cliffs, NJ},
}

@article{marple1999computing,
  author = {Marple, S. Lawrence},
  title = {Computing the Discrete-Time ``Analytic'' Signal via {FFT}},
  journal = {IEEE Transactions on Signal Processing},
  volume = {47},
  number = {9},
  pages = {2600--2603},
  year = {1999},
  doi = {10.1109/78.782222},
}

@misc{brochu2010tutorial,
  title={A Tutorial on Bayesian Optimization of Expensive Cost Functions, with Application to Active User Modeling and Hierarchical Reinforcement Learning},
  author={Brochu, Eric and Cora, Vlad M. and de Freitas, Nando},
  year={2010},
  eprint={1012.2599},
  archivePrefix={arXiv},
  primaryClass={cs.LG},
  doi={10.48550/arXiv.1012.2599}
}

@article{hjorth1975online,
  author = {Hjorth, B.},
  title = {An On-line Transformation of {EEG} Scalp Potentials into Orthogonal Source Derivations},
  journal = {Electroencephalography and Clinical Neurophysiology},
  volume = {39},
  number = {5},
  pages = {526--530},
  year = {1975},
  doi = {10.1016/0013-4694(75)90056-5}
}

@inproceedings{zeng2023transformers,
  author = {Zeng, Ailing and Chen, Muxi and Zhang, Lei and Xu, Qiang},
  title = {Are Transformers Effective for Time Series Forecasting?},
  booktitle = {Proceedings of the AAAI Conference on Artificial Intelligence},
  volume = {37},
  pages = {11121--11128},
  year = {2023},
  doi = {10.1609/aaai.v37i9.26317}
}

@article{zrenner2023rising,
  author = {Zrenner, Christoph and Koz{\'a}k, G{\'a}bor and Schaworonkow, Natalie and Metsomaa, Johanna and Baur, David and Vetter, Dragana and others},
  title = {Corticospinal Excitability Is Highest at the Early Rising Phase of Sensorimotor {$\mu$}-rhythm},
  journal = {NeuroImage},
  volume = {266},
  pages = {119805},
  year = {2023},
  doi = {10.1016/j.neuroimage.2022.119805}
}

@article{zrenner2016twoloops,
  author = {Zrenner, Christoph and Belardinelli, Paolo and M{\"u}ller-Dahlhaus, Florian and Ziemann, Ulf},
  title = {Closed-Loop Neuroscience and Non-invasive Brain Stimulation: {A} Tale of Two Loops},
  journal = {Frontiers in Cellular Neuroscience},
  volume = {10},
  pages = {92},
  year = {2016},
  doi = {10.3389/fncel.2016.00092}
}

@article{figee2021personalized,
  author = {Figee, Martijn and Mayberg, Helen},
  title = {The Future of Personalized Brain Stimulation},
  journal = {Nature Medicine},
  volume = {27},
  number = {2},
  pages = {196--197},
  year = {2021},
  doi = {10.1038/s41591-021-01243-7}
}

@article{article,
author = {Chang, Yu-Cheng and Chao, Pin-Hsuan and Kuan, Yan-Ming and Huang, Chiu-Jung and Chen, Li-Fen and Mao, Wei-Chung and Su, Tzu and Chen, Sin-Horng and Wei, Chun-Shu},
year = {2025},
title = {Forecast Padding Enhances Accuracy and Robustness of EEG-Phase-Synchronized TMS},
journal = {IEEE transactions on neural systems and rehabilitation engineering : a publication of the IEEE Engineering in Medicine and Biology Society},
doi = {10.1109/TNSRE.2025.3587711}
}

@article{babayan2019mind,
  title={A mind-brain-body dataset of MRI, EEG, cognition, emotion, and peripheral physiology in young and old adults},
  author={Babayan, Anahit and Erbey, Miray and Kumral, Deniz and Reinelt, Janis D and Reiter, Andrea MF and R{\"o}bbig, Josefin and Schaare, H Lina and Uhlig, Marie and Anwander, Alfred and Bazin, Pierre-Louis and others},
  journal={Scientific Data},
  volume={6},
  number={1},
  pages={180308},
  year={2019},
  doi={10.1038/sdata.2018.308}
}

@article{mcfarland1997spatial,
  author = {McFarland, Dennis J. and McCane, Lynn M. and David, Stephen V. and Wolpaw, Jonathan R.},
  title = {Spatial filter selection for {EEG}-based communication},
  journal = {Electroencephalography and Clinical Neurophysiology},
  year = {1997},
  volume = {103},
  number = {3},
  pages = {386--394},
  doi = {10.1016/S0013-4694(97)00022-2},
}

@misc{kirchhoff2026phaseestimationautoregressivepadding,
    title={Phase estimation with autoregressive padding (PEAP): addressing inaccuracies and biases in EEG analysis}, 
    author={Miriam Kirchhoff and Johanna Rösch and Maria Ermolova and Oskari Ahola and Sarah Harders and Juliana Hougland and Ulf Ziemann},
    year={2026},
    eprint={2604.02212},
    archivePrefix={arXiv},
    primaryClass={q-bio.NC},
}

@article{Gustafsson1996,
  author = {Gustafsson, Fredrik},
  title = {Determining the Initial States in Forward-Backward Filtering},
  journal = {IEEE Transactions on Signal Processing},
  year = {1996},
  volume = {44},
  number = {4},
  pages = {988--992},
  month = apr,
  doi = {10.1109/78.492552},
}

@article{mansouri2017fast,
  author = {Mansouri, Farrokh and Dunlop, Katharine and Giacobbe, Peter and Downar, Jonathan and Zariffa, Jos{\'e}},
  title = {A Fast {EEG} Forecasting Algorithm for Phase-Locked Transcranial Electrical Stimulation of the Human Brain},
  journal = {Frontiers in Neuroscience}, year = {2017}, volume = {11}, pages = {401},
  doi = {10.3389/fnins.2017.00401},
}

@article{gordon2021prefrontal,
  author = {Gordon, Pedro Caldana and D{\"o}rre, Sara and Belardinelli, Paolo and Stenroos, Matti and Zrenner, Brigitte and Ziemann, Ulf and Zrenner, Christoph},
  title = {Prefrontal Theta-Phase Synchronized Brain Stimulation With Real-Time {EEG}-Triggered {TMS}},
  journal = {Frontiers in Human Neuroscience}, year = {2021}, volume = {15}, pages = {691821},
  doi = {10.3389/fnhum.2021.691821},
}

@article{butterworth1930,
  author = {Butterworth, Stephen},
  title = {On the Theory of Filter Amplifiers},
  journal = {Experimental Wireless and the Wireless Engineer},
  year = {1930}, volume = {7}, pages = {536--541}, month = oct,
}

@book{oppenheim2010,
  author = {Oppenheim, Alan V. and Schafer, Ronald W.},
  title = {Discrete-Time Signal Processing},
  edition = {3rd}, year = {2010},
  publisher = {Prentice Hall}, address = {Upper Saddle River, NJ},
  isbn = {978-0131988422},
}

@article{levinson1947wiener,
  author = {Levinson, Norman},
  title = {The {W}iener {RMS} (Root Mean Square) Error Criterion in Filter Design and Prediction},
  journal = {Journal of Mathematics and Physics},
  year = {1947},
  volume = {25},
  number = {1--4},
  pages = {261--278},
  doi = {10.1002/sapm1946251261}
}

@article{durbin1960fitting,
  author = {Durbin, James},
  title = {The Fitting of Time-Series Models},
  journal = {Revue de l'Institut International de Statistique / Review of the International Statistical Institute},
  year = {1960},
  volume = {28},
  number = {3},
  pages = {233--243},
  doi = {10.2307/1401322}
}

@article{trench1964algorithm,
  author = {Trench, William F.},
  title = {An Algorithm for the Inversion of Finite {T}oeplitz Matrices},
  journal = {Journal of the Society for Industrial and Applied Mathematics},
  year = {1964},
  volume = {12},
  number = {3},
  pages = {515--522},
  doi = {10.1137/0112045}
}

@techreport{musicus1988levinson,
  author = {Musicus, Bruce R.},
  title = {Levinson and Fast {C}holeski Algorithms for {T}oeplitz and Almost {T}oeplitz Matrices},
  institution = {Research Laboratory of Electronics, Massachusetts Institute of Technology},
  year = {1988},
  number = {538},
  type = {RLE Technical Report},
  address = {Cambridge, MA},
}

@article{delsarte1986split,
  author = {Delsarte, Philippe and Genin, Yves V.},
  title = {The Split {L}evinson Algorithm},
  journal = {IEEE Transactions on Acoustics, Speech, and Signal Processing},
  year = {1986},
  volume = {34},
  number = {3},
  pages = {470--478},
  doi = {10.1109/TASSP.1986.1164830}
}

@book{mills1990timeseries,
  author = {Mills, Terence C.},
  title = {Time Series Techniques for Economists},
  publisher = {Cambridge University Press},
  year = {1990},
  address = {Cambridge},
  isbn = {9780521343398}
}

@book{percival1993spectral,
  author = {Percival, Donald B. and Walden, Andrew T.},
  title = {Spectral Analysis for Physical Applications},
  publisher = {Cambridge University Press},
  year = {1993},
  address = {Cambridge},
  isbn = {9780521435413},
  doi = {10.1017/CBO9780511622762}
}

@book{pandit1983timeseries,
  author = {Pandit, Sudhakar M. and Wu, Shien-Ming},
  title = {Time Series and System Analysis with Applications},
  publisher = {John Wiley \& Sons},
  year = {1983},
  address = {New York},
  isbn = {9780471867364}
}

@inproceedings{akiba2019optuna,
  title = {Optuna: A Next-generation Hyperparameter Optimization Framework},
  author = {Akiba, Takuya and Sano, Shotaro and Yanase, Toshihiko and Ohta, Takeru and Koyama, Masanori},
  booktitle = {Proceedings of the 25th ACM SIGKDD International Conference on Knowledge Discovery \& Data Mining},
  pages = {2623--2631},
  year = {2019},
  publisher = {ACM},
  doi = {10.1145/3292500.3330701}
}

@article{bergstra2012random,
  title = {Random Search for Hyper-Parameter Optimization},
  author = {Bergstra, James and Bengio, Yoshua},
  journal = {Journal of Machine Learning Research},
  volume = {13},
  number = {10},
  pages = {281--305},
  year = {2012}
}

@article{DBCNNHamdan,
  author = {Hamdan, Emadeldeen and Luo, Yingyi and Forelli, Ryan and Liufu, Mengzhan and Zhou, Nan and Shridhar, Sameera and Quattrocchi, Ellie and Leveroni, Zachary and Ogrenci, Seda and Tran, Nhan and Cetin, Ahmet Enis and Yu, Jai Y.},
  journal = {IEEE Transactions on Biomedical Engineering}, 
  title = {Real-Time Instantaneous Phase Estimation Using a Deep Dual-Branch Complex Neural Network}, 
  year = {2025},
  volume = {73},
  number = {9},
  pages = {3128-3139},
  doi = {10.1109/TBME.2025.3647598}}

@article{vanengen2026dissociating,
  author = {van Engen, Q. and Chau, G. and Smith, A. and Adam, K. and Donoghue, T. and Voytek, B.},
  year = {2026},
  title = {Dissociating Contributions of Theta and Alpha Oscillations from Aperiodic Neural Activity in Human Visual Working Memory},
  journal = {Journal of Neuroscience},
  volume = {46},
  number = {7},
  pages = {e2340242025},
  doi = {10.1523/JNEUROSCI.2340-24.2025}
}

@article{donoghue2022methodological,
  author = {Donoghue, T. and Schaworonkow, N. and Voytek, B.},
  year = {2022},
  title = {Methodological considerations for studying neural oscillations},
  journal = {European Journal of Neuroscience},
  volume = {55},
  number = {11--12},
  pages = {3502--3527},
  doi = {10.1111/ejn.15361}
}

@inproceedings{blackwood2018continuous,
  author = {Blackwood, Ethan and Lo, Meng-chen and Widge, Alik S.},
  title = {Continuous Phase Estimation for Phase-Locked Neural Stimulation Using an Autoregressive Model for Signal Prediction},
  booktitle = {2018 40th Annual International Conference of the IEEE Engineering in Medicine and Biology Society (EMBC)},
  year = {2018},
  pages = {4736--4739},
  address = {Honolulu, HI},
  doi = {10.1109/EMBC.2018.8513232}
}

@article{wodeyar2023different,
  author = {Wodeyar, Anirudh and Marshall, Fran{\c{c}}ois A. and Chu, Catherine J. and Eden, Uri T. and Kramer, Mark A.},
  title = {Different Methods to Estimate the Phase of Neural Rhythms Agree But Only During Times of Low Uncertainty},
  journal = {eNeuro},
  year = {2023},
  volume = {10},
  number = {11},
  pages = {ENEURO.0507-22.2023},
  doi = {10.1523/ENEURO.0507-22.2023}
}

@article{sobol2001global,
  author = {Sobol', Ilya M.},
  title = {Global Sensitivity Indices for Nonlinear Mathematical Models and Their {M}onte {C}arlo Estimates},
  journal = {Mathematics and Computers in Simulation},
  volume = {55}, number = {1--3}, pages = {271--280}, year = {2001},
  doi = {10.1016/S0378-4754(00)00270-6}
}

@article{wilcoxon1945individual,
  author = {Wilcoxon, Frank},
  title = {Individual Comparisons by Ranking Methods},
  journal = {Biometrics Bulletin},
  volume = {1}, number = {6}, pages = {80--83}, year = {1945},
  doi = {10.2307/3001968}
}

@article{welch1947generalization,
  author = {Welch, Bernard L.},
  title = {The Generalization of ``{S}tudent's'' Problem when Several Different Population Variances Are Involved},
  journal = {Biometrika},
  volume = {34}, number = {1--2}, pages = {28--35}, year = {1947},
  doi = {10.1093/biomet/34.1-2.28}
}

@article{fisher1915frequency,
  author = {Fisher, Ronald A.},
  title = {Frequency Distribution of the Values of the Correlation Coefficient in Samples from an Indefinitely Large Population},
  journal = {Biometrika},
  volume = {10}, number = {4}, pages = {507--521}, year = {1915},
  doi = {10.2307/2331838}
}

@article{holm1979simple,
  author = {Holm, Sture},
  title = {A Simple Sequentially Rejective Multiple Test Procedure},
  journal = {Scandinavian Journal of Statistics},
  volume = {6}, number = {2}, pages = {65--70}, year = {1979}
}

@book{efron1993introduction,
  author = {Efron, Bradley and Tibshirani, Robert J.},
  title = {An Introduction to the Bootstrap},
  publisher = {Chapman \& Hall}, address = {New York}, year = {1993},
  doi = {10.1201/9780429246593}
}

@inproceedings{bergstra2011algorithms,
  author = {Bergstra, James and Bardenet, R{\'e}mi and Bengio, Yoshua and K{\'e}gl, Bal{\'a}zs},
  title = {Algorithms for Hyper-Parameter Optimization},
  booktitle = {Advances in Neural Information Processing Systems},
  volume = {24}, pages = {2546--2554}, year = {2011}
}

\end{document}